\documentclass[12pt]{article}
\usepackage[english]{babel}
\usepackage{amsfonts}
\usepackage{amssymb}
\usepackage{latexsym}
\usepackage{graphicx}
\usepackage{epsfig}
\usepackage{epsf}

\newcommand{\sectiono}[1]{\section{#1}\setcounter{equation}{0}}

\newcommand{\be}{\begin{equation}}
\newcommand{\ee}{\end{equation}}
\newcommand{\bi}{\begin{itemize}}
\newcommand{\ei}{\end{itemize}}
\newcommand{\bea}{\begin{eqnarray}}
\newcommand{\eea}{\end{eqnarray}}
\newcommand{\ba}{\begin{array}}
\newcommand{\ea}{\end{array}}
\newcommand{\nn}{\nonumber}
\begin{document}
\begin{titlepage}
\rightline{GEF-TH-05-2009}
\vskip 1in
\begin{center}
\def\thefootnote{\fnsymbol{footnote}}
{\large \bf The Coupling of Chern-Simons Theory to Topological Gravity}
\vskip 0.3in
Camillo Imbimbo\footnote{E-Mail: camillo.imbimbo@ge.infn.it} 
\vskip .2in
{\em Dipartimento di Fisica, Universit\`a di Genova\\
and\\ Istituto Nazionale di Fisica Nucleare, Sezione di Genova\\
via Dodecaneso 33, I-16146, Genoa, Italy}
\end{center}
\vskip .4in
\begin{abstract}

We couple Chern-Simons gauge theory to 3-dimensional topological gravity
with the aim of investigating its quantum topological invariance.
We derive the relevant BRST rules and Batalin-Vilkovisky action.
Standard BRST transformations of the gauge field are modified by
terms involving both its anti-field and the super-ghost of topological
gravity.  Beyond the obvious couplings to the metric and the
gravitino, the BV action includes hitherto neglected couplings to the
super-ghost.  We use this result to determine the topological
anomalies of certain higher ghost deformations of $SU(N)$ Chern-Simons
theory, introduced years ago by Witten.  In the context of topological
strings these anomalies, which generalize the familiar framing
anomaly, are expected to be cancelled by couplings of the closed
string sector.  We show that such couplings are obtained by dressing
the closed string field with topological gravity observables.

\end{abstract}
\vfill
\end{titlepage}

\setcounter{footnote}{0}

\sectiono{Introduction and  Summary}

Classical Chern-Simons (CS) theory \cite{Schwarz:1978cn} on a 3-dimensional manifold $M_3$ 
is both gauge invariant and invariant under space-time diffeomorphisms,
without the need to introduce a space-time metric $g_{\mu\nu}$. 
In the quantum theory the metric appears in the gauge-fixing term:
The issue regarding topological anomalies is whether or not quantum averages 
depend on the chosen metric.

It was understood by Witten,
in his celebrated paper \cite{Witten:1988hf}, 
that quantum CS theory indeed suffers from a topological anomaly, 
the so-called {\it framing} anomaly\footnote{For a discussion which expands 
on the local presentation of the framing anomaly, the point of view of
the present paper, see also \cite{BarNatan:1991rn}.}. In the present work we 
address the question of quantum topological invariance 
in certain generalizations of CS gauge theory which
were introduced by Witten in \cite{Witten:1992fb} and  which are
obtained by adding observables with higher ghost number to the classical CS action.

We will formulate the problem by making use of a trick
which goes back to the early days of BRST renormalization methods\footnote{This idea,
which appears to have been known to BRST experts for quite a long time,  was 
rediscovered several times in different contexts. It was applied, for example, in \cite{KlubergStern:1974rs}  to Yang-Mills theory and in
\cite{Piguet:1984mv} to supersymmetry.} and which
can be explained as follows. Consider a gauge-fixed action
\be
\Gamma(\alpha^i) = \Gamma_0 + S_0\,\chi(\alpha^i)
\ee
where $\Gamma_0$ is the classical action, $S_0$ is the BRST operator, $\chi(\alpha^i)$
is the gauge-fermion which depends on  the commuting 
parameters $\alpha^i$. To simplify the notation we dropped any references 
to either fields or sources. The goal is
understanding the (in)dependence on $\alpha^i$ of quantum averages,
which we schematically denote as 
\be
Z(\alpha^i) = \int {\rm e}^{\frac{i}{\hbar}\Gamma (\alpha^i)}
\label{parametriczeta}
\ee
To study this question one extends the action of $S_0$ to the
parameters $\alpha^i$ by defining a new nilpotent BRST operator $\hat{S}$
\be
\hat{S} \equiv S_0 + \beta^i\partial_{\alpha^i}
\label{extendeds}
\ee
where $\beta^i$ are anti-commuting variables. The $\hat{S}$-invariant classical 
action
\be
\hat{\Gamma}(\alpha^i, \beta^i) \equiv \Gamma_0 + \hat{S}\,\chi(\alpha^i)
\label{actionextended}
\ee
defines a new ``partition function'' which, in general, depends on both $\alpha^i$ and $\beta^i$:
\be
Z(\alpha^i, \beta^i) = \int {\rm e}^{\frac{i}{\hbar}\hat{\Gamma} (\alpha^i,\beta^i)}
\ee
$Z(\alpha^i, \beta^i)$ satisfies, {\it up to anomalies}, the identity
\be
\hat{S} \, Z(\alpha^i, \beta^i) =0
\label{shatinvariance}
\ee 
It admits an expansion in terms of the anti-commuting variables
\be
Z(\alpha^i, \beta^i)= Z^{(0)}(\alpha^i)+ \beta^i\, Z_i^{(1)}(\alpha^i)+
\beta^i\,\beta^j Z_{ij}^{(2)}(\alpha^i)+\cdots
\ee
whose first term is the original quantum average one is interested in
\be
Z^{(0)}(\alpha^i)=Z(\alpha^i)
\ee
The identity (\ref{shatinvariance}) translates into identities for
each one of the  $Z^{(k)}_{i_1,\ldots,i_k}(\alpha^i)$.  The first one, 
with $k=0$, is:
\be
\beta^i \frac{\partial}{\partial\alpha_i}\, Z^{(0)}(\alpha^i)=0 \Rightarrow
 \frac{\partial}{\partial\alpha_i}\, Z(\alpha^i)=0
\ee
It just states the independence of the original $Z(\alpha^i)$ from
the gauge parameters $\alpha^i$. The advantage of reformulating the (classical)
gauge independence of  $Z(\alpha^i)$ as the cocyle condition 
(\ref{shatinvariance}) for its extension $Z(\alpha^i, \beta^i)$ is that this
permits the use of  powerful cohomological methods to investigate the corresponding quantum anomalies.

In traditional quantum field theory the introduction of 
$Z(\alpha^i, \beta^i)$ is just a technical trick: The
components $Z^{(k)}_{i_1,\ldots i_k}(\alpha^i)$ with $k>0$ have usually
no physical interest. As a matter of fact, they typically  vanish, due to
ghost number conservation,  since the $\beta^i$'s carry ghost number +1.
Both in string theory and in topological field theories the situation
is different since observables have generally 
non-vanishing positive ghost number. The classical
identity (\ref{shatinvariance}) says that the components of fixed 
$\beta$-degree of the generalized partition function 
\be
Z^{(k)}(\alpha^i, \beta^i) \equiv\beta^{i_1}\,\cdots\beta^{i_k}\,Z^{(k)}_{i_1\cdots i_{k}}(\alpha^i)
\ee 
 should be thought of as {\it closed} $k$-forms on
the manifold parameterized by the $\alpha^i$:
\bea
&&\beta^{i_1}\,\beta^{i_2}\,\cdots\beta^{i_k}\,\beta^{i_{k+1}}\partial_{i_1}\,
Z^{(k)}_{i_2\cdots i_{k+1}}(\alpha^i) =0
\eea
In string theory, for example, the parameters $\alpha^i$ are coordinates on the
moduli space of Riemann surfaces  and the total ghost number of the observables is such
that the non-vanishing component of $Z(\alpha, \beta)$ is a top-form
on the relevant moduli space \cite{Becchi:1995ik}.

In the context of CS theory, the role of the
parameters $\alpha^i$ is played by the background metric $g_{\mu\nu}(x)$. 
The BRST operator $S_0$ is the one encoding gauge invariance of the classical
action.
It was therefore suggested by Witten in \cite{Witten:1992fb} to introduce
a fermionic gravitino background $\psi_{\mu\nu}(x)$ and, in analogy with (\ref{extendeds}),  to extend the action
of the BRST operator to the backgrounds
\bea
&&\hat{S}\, g_{\mu\nu} = \psi_{\mu\nu} \qquad \hat{S}\,\psi_{\mu\nu}=0
\label{brszero}
\eea
In the same paper Witten proposed also to add to the CS
classical action observables with positive ghost number. In this way one would
obtain non-vanishing components $Z^{(k)}$ with $k>0$ which
would naturally define cohomology classes on the space of 
3-dimensional metrics on the manifold $M_3$. 

In this paper we reconsider and refine this proposal by starting from the following observation. The BRST rules (\ref{brszero}) are those of topological gravity 
\cite{Witten:1988xi}. Or, more precisely, they are the {\it ``naive''}
BRST transformations of topological gravity. The correct ones are 
the {\it equivariant} ones
\cite{Baulieu:1988xs,Ouvry:1988mm,Baulieu:1989rs,Kanno:1988wm}
\bea
&& s \,g_{\mu\nu} = \psi_{\mu\nu} -{\cal L}_\xi\, g_{\mu\nu}\qquad
s \,\psi_{\mu\nu} = {\cal L}_\gamma\, g_{\mu\nu}-{\cal L}_\xi\, 
\psi_{\mu\nu}\nonumber\\
&&s \,\xi^\mu = \gamma^\mu -\frac{1}{2}\, {\cal L}_\xi\, \xi^\mu\qquad
 s \,\gamma^{\mu} =-{\cal L}_\xi\,\gamma^\mu
\label{brsgravityequivariantintro}
\eea
which involve, beyond the metric and the gravitino, the reparametrization
ghost $\xi^\mu$, the corresponding super-ghost\footnote{$\gamma^\mu$ 
can also be thought of as a ``ghost-for-ghost''.} $\gamma^\mu$ (of ghost number +2) and infinitesimal
diffeomorphisms ${\cal L}_\xi$ (${\cal L}_\gamma$) with parameters $\xi^\mu$ ($\gamma^\mu$). The reason to go from (\ref{brszero}) to (\ref{brsgravityequivariantintro}) is 
to ensure that the closed forms $Z^{(k)}$  descend to globally defined forms on the quotient of the space of  3-dimensional metrics under the action of diffeomorphisms\cite{Ouvry:1988mm,Becchi:1995ik}. This will be reviewed in Section \ref{effectiveaction}.

We find that once we extend the action of the BRST operator to the
gravitational backgrounds in the equivariant way, we must --- to
preserve nilpotency --- modify the BRST transformations in the gauge
sector.  We will work this out in Section
\ref{cstopologicalstructure}.  It turns out that the new BRST operator
can be expressed in terms of an operator $S$ which writes as follows
\be
S\equiv \hat{S}+G_\gamma
\ee
Here $\hat{S}$ is the ``naive'' extension (\ref{brszero})  
of the standard BRST operator of gauge theories to the
gravitational backgrounds.  $G_\gamma$ is the novel term dictated by equivariance which
is at the center of our analysis. It has degree 1 in the
super-ghost field $\gamma^\mu$ and it deforms in a non-trivial way the
BRST rules of the gauge sector.  $\hat{S}$ and $G_\gamma$ satisfy an
$N=2$ supersymmetry algebra
\be
\hat{S}^2= G_\gamma^2 =0\qquad  \{\hat{S}, G_\gamma\} ={\cal L}_\gamma
\ee
and the complete 
nilpotent equivariant BRST operator is simply $s=S- {\cal L}_\xi$.

An important point is that the $G_\gamma$-transformation of the gauge fields
depends on their anti-fields. In other words, a proper off-shell formulation
of  $G_\gamma$ requires adopting the Batalin-Vilkovisky formalism.

Having deformed the BRST operator in a non-trivial way, one needs to reconsider
the classical action. This is not simply given by the obvious analog of
(\ref{actionextended}), precisely because the
BRST variation of the gauge field contains its anti-field. 
We present the complete Batalin-Vilkovisky classical action
in Section \ref{bvaction}. The peculiar feature of this action is a term which is
{\it quadratic} in the anti-fields and linear in the super-ghost field
$\gamma^\mu$. We believe this term is new and has not been worked out
before. 

The somewhat surprising outcome of this analysis is that the consistent
formulation of CS theory on a curved manifold  entails
coupling it not only to the metric and the gravitino backgrounds, but also 
to the super-ghost field\footnote{The reparametrization ghost $\xi^\mu$
drops out of the gauge-fixed action, as dictated by  equivariance.}.
We make a digression in Section \ref{globalvectorsusy} to 
understand this from the point of view of the theory in flat space.
In essence, the story goes as follows. The theory in flat space has a stress-energy tensor $T_{\mu\nu}$ which is BRST-exact
\be
T_{\mu\nu}= S_0\, G_{\mu\nu}
\ee
$G_{\mu\nu}$ is defined by this relation only up to $S_0$-exact terms. 
It turns out that if one takes it  to be {\it symmetric} with respect to the exchange of $\mu$ and $\nu$, as appropriate for the current coupled to the gravitino, 
$G_{\mu\nu}$ is conserved only up to $S_0$-trivial terms. One can define a truly conserved super-current $\tilde{G}_{\mu\nu}$, but this
necessarily has an anti-symmetric piece: While the symmetric part of  $\tilde{G}_{\mu\nu}$
couples to the
gravitino, the anti-symmetric part needs  a
vector field to couple to.  This explains the necessity of the source $\gamma^\mu$
from the point of view of currents  in flat space. 
Incidentally,  the conserved $\tilde{G}_{\mu\nu}$, which is the remnant in flat space of the BRST charge $G_\gamma$,  is the super-current associated
to the vector super-symmetry of gauge-fixed CS in flat space
which was discovered years ago \cite{Delduc:1989ft}. The coupling
to topological gravity provides, in a sense, a more conceptual explanation of this flat space global symmetry.

In Section \ref{higherghostCS} we reconsider, in the equivariant
context, the observables of positive ghost number put forward by
Witten. We find that one can build {\it equivariant} gauge observables
if one includes the anti-fields of the gauge sector.  Since the
observables involve anti-fields, adding them to the theory will, in
general, not only modify the action but also deform the BRST
transformations. We work out in some detail the case of gauge group
$G=SU(N)$. For this theory we find $N-1$ ``primitive'' single trace
observables of positive ghost number, all other observables being
multi-traces of the primitive ones. The single-trace higher-ghost 
deformation of $SU(N)$ CS theory depends therefore on $N-1$ parameters
$t_i$, with $i=1,\ldots N-1$, in addition to the gauge coupling constant. 
The corresponding $Z^{(2\,p)}$ are
$2\,p$-forms on the moduli space of 3-dimensional metrics, which are
homogeneous polynomials of degree $p$ in the $t_i$'s , with $t_i$ of
weight $i$.

In Section \ref{topanomalies} we discuss at last the topological anomalies 
of the deformed CS theory, that is the failure of the  forms $Z^{(2\,p)}$ to be closed. Here the trick (\ref{extendeds}) pays off, since the problem
is reduced to determining local observables of 3-dimensional topological
gravity. Using methods \cite{Myers:1989dn,Myers:1990zi}
which have been applied in the literature mostly in dimension 2 and 4, 
we find that topological gravity observables in 3d --- and thus CS topological anomalies --- are powers of a single basic observable
\be
{\cal A}_{4\,p} = \bigl({\rm tr} {\cal R}^2\bigr)^{p}\qquad p=1,2,\ldots
\ee
${\cal R}$ is a generalized curvature form of total 
(form+ghost number) fermionic degree 2, 
whose 2-form component is the curvature of the
background metric $g_{\mu\nu}$. 
The $p=1$ observable is Witten's framing anomaly, the $p>1$ observables are 
higher-ghost generalizations of it. 

The cohomological analysis does not yield the ``coefficients'' of the anomalies.
These ``coefficients'' are in fact polynomials $c_{2\,(p-1)}(t_i)$ 
in $t_i$, homogeneous of degree $2\,(p-1)$, with $t_i$ of weight $i$. 
It is possible that the numerical coefficients of such polynomials have 
some interesting  topological interpretation. 
We postpone to the future attempts to 
compute them explicitly.

In the last Section \ref{topinflow} we try to integrate our analysis
with the insight of Witten \cite{Witten:1992fb} that 
$SU(N)$ CS theory on $M_3$ is the target space 
field theory of topological open strings propagating on the non-compact
Calabi-Yau (complex) 3-manifold $X_6=T^* M_3$. 
One expects that the complete closed and
open topological string theory be anomaly free. We propose
a simple and natural way to introduce couplings in the closed string
field theory which cancel the higher-ghost anomalies of the deformed
$SU(N)$ CS theory. Following  \cite{Bershadsky:1994sr} we 
assume that  the closed string
field ${\cal K}$ be a generalized form of total (form +
ghost number) degree equal to 2. Then the coupling of the closed string field to
the observables of topological gravity
\be
\Gamma_{anomaly} = \sum_p  c_{2\,(p-1)}(t_i)\int_{X_6} {\cal K}\,\bigl({\rm tr} {\cal R}^2\bigr)^{p}
\ee
cancel the topological anomalies of the open string sector. In \cite{Gopakumar:1998ki}
it was  checked essentially that the $p=1$ term of this sum is indeed consistent with what is known about the partition function of the topological closed string model. 

Beyond the explicit computation of the anomaly polynomials $c_{2\,(p-1)}(t_i)$ 
there are several other problems that are left open by this work. 
The relation between the anomalies of $SU(N)$ deformed CS and 
6-dimensional topological gravity is somewhat reminiscent of the 
Kontsevitch relation between large  $N$ matrix theory and 2-dimensional 
topological gravity \cite{Kontsevich:1992ti},\cite{Gaiotto:2003yb}. 
One might speculate if this analogy could lead to a combinatorial interpretation
of the anomaly  polynomials $c_{2\,(p-1)}(t_i)$ and to an efficient way to compute them.

It would be also important to understand how one computes, on the topological string side, amplitudes involving CS couplings with higher ghost number. We believe that
playing the trick (\ref{extendeds}) at the level of the topological sigma model maybe useful 
in this regard. But the details have not been worked out. Understanding this issue
might open the way to significant and possibly fruitful generalizations of the open-closed topological string duality of \cite{Gopakumar:1998ki}.

\sectiono{Chern-Simons Topological Structure}
\label{cstopologicalstructure}

Let
\be
A = A^a_\mu\, T^a\, dx^\mu
\ee
be a 1-form gauge field on a closed 3-manifold $M_3$. $T^a$, with
 $a=1,\ldots, {\rm dim}\, G$, are generators of
the Lie algebra of the gauge group $G$
which will be taken to be simple, connected and simply connected. 
The classical CS action \cite{Schwarz:1978cn}
 writes as follows
\be
\Gamma_{CS} =  \int_{M_3} {\rm Tr}\Big[ \frac{1}{ 2}\,A\,d\,A +\frac{1}{ 3} A^3\Bigr]
\label{classicalaction}
\ee
Gauge invariance leads to the nilpotent BRST transformation rules\footnote{We will adopt the convention that BRST operator $S_0$ {\it anti-commutes} with
the exterior differential $d$.}
\bea
&& S_0 \, A = D\, c\qquad S_0\, c = c^2
\label{brszerofields}
\eea
where $c = c^a\, T^a$ is the ghost field carrying ghost number $+1$ and
$D\, c\equiv d\,c +[A, c]$ is the covariant differential.

It is useful to associate to $A$ and $c$,  respectively, the anti-fields $A^*$ and $c^*$, of ghost number $-1$ and $-2$. $A^*$ and $c^*$  are  
Lie algebra-valued 2 and 3-forms:
\bea
&&A^* = (A^*)^a_{\mu\nu}\, T^a\, dx^\mu\,dx^\nu\qquad 
c^* = (c^*)^a_{\mu\nu\rho}\, T^a\, dx^\mu\,dx^\nu\, dx^\rho
\eea
The corresponding Batalin-Vilkovisky action
\be
\Gamma_{BV} = \Gamma_{CS}[A] + \int_{M_3}{\rm Tr}\Bigl[\bigl(S_0\,A\bigr)\, A^* +\bigl(S_0 \,c\bigr)\, c^*\Bigr]\nn
\ee
is $S_0$ invariant if one extends the action of $S_0$ to the anti-fields 
in the following way
\bea
&& S_0 \, A^* = F + [A^*,c] \qquad  S_0\, c^* = D\,A^* +[c^*,c]
\label{brszeroantifields}
\eea
where the square brackets denote either commutator or anti-commutator according to the ghost numbers of the fields involved.
The BRST operator $S_0$ defined by (\ref{brszerofields}) and (\ref{brszeroantifields}) is  nilpotent on both the fields and the anti-fields. 

The classical action (\ref{classicalaction}) is invariant not just 
under diffeomorphisms
of $M_3$: it is also {\it topological}, i.e.  it is independent of the 3-dimensional background
metric $g_{\mu\nu}$. The gauge-fixed action necessarily depends on $g_{\mu\nu}$.
The issue of topological anomalies is if quantum averages do depend
on $g_{\mu\nu}$ or not.

To discuss this issue one extends the action of $S_0$ to the
background metric\footnote{To reduce symbols proliferation, we denote
the extended BRST operator with the same symbol $S_0$ as the original
one. In the Introduction we used the symbol $\hat{S}$ for the same object.}
\be
S_0 \,g_{\mu\nu} = \psi_{\mu\nu}\qquad  S_0 \,\psi_{\mu\nu} =0
\label{brsgravity}
\ee
where $\psi_{\mu\nu}$ is the topological fermionic gravitino field with ghost number
$+1$. Such an extension of the BRST transformations for CS theory
was first considered in \cite{Witten:1992fb}.

The BRST transformation rules (\ref{brsgravity}) are essentially 
those of topological gravity as it was originally defined in 
\cite{Witten:1988xi}. In this sense, to discuss
topological invariance of CS gauge theory one needs 
to couple it to topological gravity. 

Since the original paper \cite{Witten:1988xi}, 
however, it has been understood \cite{Baulieu:1988xs,Ouvry:1988mm,Baulieu:1989rs,Kanno:1988wm} that the ``naive''  BRST
rules (\ref{brsgravity}) must be slightly modified\footnote{We review the reason for this in Section \ref{effectiveaction}.}.
The correct BRST operator of topological gravity is the {\it equivariant} one,
which, in some sense, factors out diffeomorphisms. 
This necessitates introducing both the anti-commuting  ghost fields
 $\xi^\mu$ of ghost number $+1$, associated to 3-dimensional diffeomorphisms,
and its super-partner, the commuting super-ghost field $\gamma^\mu$ of ghost
number $+2$. Eqs. (\ref{brsgravity}) are to be replaced by the  
equivariant  ones
\bea
&& s \,g_{\mu\nu} = \psi_{\mu\nu} -{\cal L}_\xi\, g_{\mu\nu}\qquad
s \,\psi_{\mu\nu} = {\cal L}_\gamma\, g_{\mu\nu}-{\cal L}_\xi\, 
\psi_{\mu\nu}\nonumber\\
&&s \,\xi^\mu = \gamma^\mu -\frac{1}{2}\, {\cal L}_\xi\, \xi^\mu\qquad
 s \,\gamma^{\mu} =-{\cal L}_\xi\,\gamma^\mu
\label{brsgravityequivariant}
\eea
where ${\cal L}_\xi$ denotes the Lie derivative along the vector field $\xi^\mu$.

Our starting and basic observation is the following. Once we switch from the ``naive'' $S_0$ to the equivariant $s$ 
in the gravitational background sector we must do the same in the ``matter'' 
gauge sector, i.e. in the sector generated by the quantum fields 
$c, A$ and by their anti-fields $c^*, A^*$.
Starting from the ansatz
\be
s = S_0 - {\cal L}_\xi+ \cdots\qquad {\rm on\; the\; gauge\; sector}
\ee
we determine the dots by requiring nilpotency 
\be
s^2=0 \qquad {\rm on\; all\; fields}
\ee 
We obtain in this way to the following  BRST action 
on the gauge sector\footnote{BRST rules which combine gauge
symmetry and topological gravity transformations had been derived, with different
methods and motivations, in
\cite{Baulieu:2005bs}. The BRST rules of \cite{Baulieu:2005bs} are quite
similar, although not identical, to the ones we present here. 
In particular the $i_\gamma(A)$ term
in the BRST variation of the $c$ ghost already makes its appearance in  
\cite{Baulieu:2005bs}. However, since the focus of that work is on higher-dimensional theories,
the details of the BRST action on gauge fields seem different 
than ours. More importantly, anti-fields are not considered in \cite{Baulieu:2005bs}. The structure of the anti-field sector
specific to 3-dimensions is essential for the off-shell nilpotency of the
transformations (\ref{brsmatterequivariant}). }
\bea
&&s\, c = c^2 - {\cal L}_\xi\, c-i_\gamma(A)\nonumber\\
&&s\, A = D\, c  - {\cal L}_\xi\, A- i_\gamma (A^*)\nonumber\\
&&s\, A^* = F- {\cal L}_\xi\, A^*- [A^*,c]-i_\gamma(c^*)\nonumber\\
&&s\,c^* =D\,A^* - {\cal L}_\xi\, c^* + [c^*,c] 
\label{brsmatterequivariant}
\eea
where $i_\gamma$ acts on forms and denotes the contraction with the vector field $\gamma^\mu$:
\be
i_\gamma(A)\equiv \gamma^\mu\,A_\mu \qquad{\rm etc.}
\ee

The relevant notion of BRST cohomology in topological gravity is the equivariant
one\footnote{See the Appendix \ref{equivariantcohomology} for a review of the relevant material.}. This means that physical observables are associated to 
the cohomology of $s$ on the algebra generated by all fields with the exclusion of the reparametrization ghosts $\xi^\mu$. 
When restricting oneself to such a space it is useful to introduce the operator
\be
S \equiv s +{\cal L}_\xi
\ee
Nilpotency of $s$ is equivalent to the following property of $S$
\be
S^2 ={\cal L}_\gamma \qquad {\rm on\; all\; fields\; but\;}\xi^\mu
\ee
Hence, $S$ is nilpotent when acting on reparametrization 
invariants functionals which are also independent of $\xi^\mu$. The action of $S$, both on quantum
fields and classical backgrounds, is 
\bea
&S\, c = c^2-i_\gamma(A)\qquad &S\, A = D\, c  - i_\gamma (A^*)\nonumber\\
&S\, A^* = F +[A^*,c]-i_\gamma(c^*)\qquad
&S\, c^* = D\,A^* +[c^*,c] \nonumber\\
& S\, g_{\mu\nu} = \psi_{\mu\nu}\qquad
&S\, \psi_{\mu\nu}= {\cal L}_\gamma\, g_{\mu\nu}\nonumber\\
& S\,\gamma^\mu =0\qquad &
\eea
$S$ can be decomposed as
\be
S= S_0 +G_\gamma
\ee
where $S_0$ is the ``naive'' nilpotent BRST operator defined in (\ref{brszerofields}), (\ref{brszeroantifields}), (\ref{brsgravity}) while $G_\gamma$ is the  nilpotent operator, linear in the field $\gamma^\mu$, defined by
\bea
&&G_\gamma\, c = -i_\gamma(A)  
\qquad G_\gamma\, A =  -i_\gamma (A^*)
\qquad G_\gamma\, A^* = -i_\gamma(c^*)
\qquad G_\gamma\, c^* = 0 \nonumber\\
&& G_\gamma\, g_{\mu\nu} = 0
\qquad G_\gamma\, \psi_{\mu\nu}= {\cal L}_\gamma\, g_{\mu\nu}
\qquad  G_\gamma\,\gamma^\mu =0 
\label{ggammamatter}
\eea 
$S_0$ and $G_\gamma$ generate a supersymmetric algebra
\be
S_0^2=0\qquad G_\gamma^2 =0\qquad \{S_0, G_\gamma\} = {\cal L}_\gamma
\ee
This $N=2$ BRST algebra exposes the topological nature of CS theory:
 Infinitesimal  diffeomorphisms are expressed as commutator of $S_0$ a second
BRST symmetry charge, $G_\gamma$.
 
\sectiono{The  BV Action}
\label{bvaction}

In this section we  construct the Batalin-Vilkovisky action\footnote{For a short review of the BV formalism, see for example \cite{Fuster:2005eg}, of which we adopt the notation.}
 associated to
the equivariant $s$. The BV action relative to the ``naive'' BRST
operator $S_0$ is {\it linear} in the anti-fields:
\be
\Gamma_0 = \Gamma_{CS}[A] + \int_{M_3}{\rm Tr}\Bigl[\bigl(S_0\,A\bigr)\, A^* +\bigl(S_0 \,c\bigr)\, c^*+ \bigl(S_0\,g_{\mu\nu}\bigr)\, (g^*)^{\mu\nu}\Bigr]
\ee 
where $(g^*)^{\mu\nu}$ is the anti-field associated to $g_{\mu\nu}$ 
transforming as a tensorial density under diffeomorphisms. This ``flat space''
BV action is equivalent to the one first presented in \cite{Axelrod:1993wr} and
also discussed in \cite{Alexandrov:1995kv}. 

To construct the BV action associated to 
the equivariant $s$ we start therefore from the analogous expression
\bea
&&\Gamma_{BV}= \Gamma_{CS} + \sum_\Phi\int_{M_3} \bigl( s\,\Phi\bigr)\, \Phi^* +\cdots\nonumber
\eea
where we denote by $\Phi$  the collection of {\it all} fields and by
$\Phi^*$ their anti-fields, both those of the gauge sector
(i.e. $A, A^*, c, c^*$) and those of the gravitational sector (i.e.  
$g_{\mu\nu},\psi_{\mu\nu},\xi^\mu,\gamma^\mu$). Explicitly, 
\bea
&&\Gamma_{BV}= \Gamma_0 - \int_{M_3}{\rm Tr}\Bigl[{\cal L}_\xi\,A\, A^* + {\cal L}_\xi\,c\, c^*+ \nonumber\\
&&\qquad + \bigl({\cal L}_\xi\,g_{\mu\nu}\bigr)\, (g^*)^{\mu\nu}+\bigl({\cal L}_\xi\,\psi_{\mu\nu}\bigr)\, (\psi^*)^{\mu\nu}+ \frac{1}{ 2}\bigl({\cal L}_\xi\,\xi^{\mu}\bigr)\, \xi^*_{\mu}+
\bigl({\cal L}_\xi\,\gamma^{\mu}\bigr)\, \gamma^*_{\mu}+\nonumber\\
&&\qquad+i_\gamma(A^*)\, A^* +i_\gamma (A)\, c^* -\bigl({\cal L}_\gamma\,g_{\mu\nu}\bigr)\, (\psi^*)^{\mu\nu}- \gamma^\mu\, \xi^*_{\mu}\Bigr]+ \cdots
\label{actionone}
\eea
where we introduced the anti-fields relative to both $\xi^\mu$ and $\gamma^\mu$.
 
However, as the dots in the r.h.s. of (\ref{actionone}) indicate, 
this action is  {\it not} invariant under $s$. This can be traced back to the term proportional to the anti-field $A^*$ in the BRST variation of $A$, giving rise
to the term $-{\rm Tr}\,i_\gamma(A^*)\, A^*$ in the action, which is 
{\it quadratic} in the anti-fields. To achieve invariance we need to half the 
coefficient of such a term; we do so by adding to the action an identical term 
with coefficient $1/2$ 
\bea
&&\Gamma_{BV}= \Gamma_{CS} + \int_{M_3}{\rm Tr}\Bigl[ 
\sum_\Phi \bigl(s\,\Phi\bigr)\, \Phi^* +\frac{1}{ 2}\, i_\gamma(A^*)\, A^*\Bigr]=\nonumber \\ 
&&\qquad =\Gamma_0 - \int_{M_3}{\rm Tr}\Bigl[\bigl({\cal L}_\xi\,A\bigr)\, A^* + \bigl({\cal L}_\xi\,c\bigr)\, c^*+ \nonumber\\
&&\qquad + \bigl({\cal L}_\xi\,g_{\mu\nu}\bigr)\, (g^*)^{\mu\nu}+\bigl({\cal L}_\xi\,\psi_{\mu\nu}\bigr)\, (\psi^*)^{\mu\nu}+ \frac{1}{ 2}\bigl({\cal L}_\xi\,\xi^{\mu}\bigr)\, \xi^*_{\mu}+
\bigl({\cal L}_\xi\,\gamma^{\mu}\bigr)\, \gamma^*_{\mu}+\nonumber\\
&&\qquad+\frac{1}{ 2}\,i_\gamma(A^*)\, A^* +i_\gamma (A)\, c^* -\bigl({\cal L}_\gamma\,g_{\mu\nu}\bigr)\, (\psi^*)^{\mu\nu}- \gamma^\mu\, \xi^*_{\mu}\Bigr]
\label{actionbv}
\eea
This is the BV action invariant  under the equivariant $s$. $\Gamma_0$ is the 
BV CS gauge action associated to the ``naive'' BRST operator $S_0$. 
The other terms in $\Gamma$ describe the couplings of CS theory to both the ghost and  the super-ghost of topological
gravity. The latter are required for the equivariance of the theory.
As far as we know, the couplings to the super-ghost $\gamma^\mu$  are new and had not been considered before.

On can check directly that the BV action (\ref{actionbv}) reproduces
the equivariant BRST transformations (\ref{brsgravityequivariant}) and 
(\ref{brsmatterequivariant}) via the known formulas\footnote{Some signs in formulas (\ref{bvone}) and (\ref{bvtwo}) are different than how they are usually written. 
This is so since in our convention the BRST operator anti-commutes with the exterior differential. With this choice, $A$, $c$  have both odd total fermion number, as well as their anti-fields and  $\Gamma_{BV}$. This convention will make formulas in Section \ref{higherghostCS} look nicer.}

\bea
&&s\, \Phi = (\Gamma_{BV}, \Phi)= \frac{\delta^R\, \Gamma_{BV}}{ \delta\,\Phi^*}
\qquad  s\, \Phi^* = (\Gamma_{BV}, \Phi^*)=\frac{\delta^R\, \Gamma_{BV}}{ \delta\,\Phi}
\label{bvone}
\eea
where we introduced the BV bracket
\bea
&&(F, G) \equiv  \sum_\Phi\frac{\delta^R\, F}{ \delta\,\Phi}\frac{\delta^L\, G}{ \delta\,\Phi^*}+\frac{\delta^R\, F}{ \delta\,\Phi^*}\frac{\delta^L\, G}{ \delta\,\Phi}
\label{bvtwo}
\eea
For example, the factor $1/2$ in front of the term in the action 
quadratic in the anti-field $A^*$  ensures that
the derivative of the action with respect to $A^*$ coincides with
the BRST transformation rule for $A$
\be
s\, A \, = \frac{\delta^R\, \Gamma_{BV}}{ \delta\,A^*} = -i_\gamma(A^*) +\cdots
\ee 

As dictated by the BV formalism, both the nilpotence of $s$ and the BRST invariance of $\Gamma$ are captured by the single equation
\be
(\Gamma_{BV}, \Gamma_{BV})=0
\ee

\subsection{The  gauge-fixed action}

Gauge-fixing is achieved by introducing a suitable functional $\chi[\Phi]$
of the fields. The gauge-fixed action $\Gamma_\chi$ is then given by the formula
\be
\Gamma_\chi[\Phi]= \Gamma_{BV}[\Phi, \Phi^*= \frac{\delta\,\chi}{ \delta\,\Phi}]
\ee
and the gauge-fixed BRST operator is  
\be
s_\chi\, \Phi = \bigl(s\,\Phi\bigr)\Big|_{ \Phi^*= \frac{\delta\,\chi}{ \delta\,\Phi}}
\ee
The square of $s_\chi$ is 
\be
s^2_\chi\, \Phi = -\sum_{\Phi^\prime} 
\frac{\delta^R\,\Gamma_\chi}{ \delta\,\Phi^\prime}\,\frac{\delta^L\,\bigl(s\,\Phi)\,}{ \delta\,(\Phi^\prime)^*}\Big|_{ \Phi^*= \frac{\delta\,\chi}{ \delta\,\Phi}}
\label{schisquare}
\ee
Therefore, when the BV action $\Gamma_{BV}[\Phi, \Phi^*]$ is not linear in the anti-fields, $s_\chi$ is nilpotent only on the (gauge-fixed) shell.

The gauge-fixed action is $s_\chi$-invariant (off-shell, of course):
\be
s_\chi\, \Gamma_\chi =0
\ee

The gauge-fermion $\chi[\Phi]$ must be chosen in such a way that all fields
have invertible
kinetic terms. To achieve this, it is typically necessary to
introduce more fields, beyond gauge and ghost (anti)fields.
For CS theory we must add both   the
anti-ghost  $b$ and the lagrangian multiplier $\Lambda$
\be
b = T^a\,b^a_{\mu\nu\rho}dx^\mu\,dx^\nu\,dx^\rho\qquad \Lambda = 
T^a\,\Lambda^a_{\mu\nu\rho}dx^\mu\,dx^\nu\,dx^\rho
\ee
which are  3-forms with values in the Lie algebra of the gauge group.
Their BRST transformation laws are 
\bea
&& s\, b = \Lambda -{\cal L}_\xi\, b\qquad s \, \Lambda = {\cal L}_\gamma\, b-{\cal L}_\xi\, \Lambda
\eea
From now on, we will denote by $\Phi$ the collection of all fields, coming from
the gauge, the gravitational and anti-ghost sectors.
 
From (\ref{actionbv}) we obtain the CS gauge-fixed action 
\be
\Gamma_\chi = \Gamma_{CS}[A] + s_\chi\,\bigl(\chi[\Phi]\bigr)-\frac{1}{ 2}\int_{M_3}\,{\rm Tr}\,\Bigl(i_\gamma\Bigl(\frac{\delta\,\chi}{ \delta\,A}\Big)\, \frac{\delta\,\chi}{ \delta\,A}\Bigr)
\label{gfaction}
\ee
and the gauge-fixed BRST transformations
\bea
&&s_\chi\, c = c^2 - {\cal L}_\xi\, c-i_\gamma(A)\nonumber\\
&&s_\chi\, A = D\, c  - {\cal L}_\xi\, A- i_\gamma (\frac{\delta\,\chi}{ \delta\,A})\label{gfbrst}
\eea
From (\ref{schisquare}) we conclude that the gauge-fixed BRST operator is 
nilpotent only up to terms proportional to the equations of motion of $A$:
\be
s_\chi^2 = -\int_{M_3}{\rm Tr}\Bigl[\,i_\gamma\Bigl(\frac{\delta\,\Gamma_\chi}{ \delta\,A}\Big)\, \frac{\delta}{ \delta\,A}\Bigr]
\ee

Note that although the gauge-fixing part of the gauge-fixed action (\ref{gfaction}) is not $s_\chi$-exact, the change of $\Gamma_\chi$ 
under variation of $\chi$ 
\be 
\chi \to \chi + \delta\chi
\ee
is $s_\chi$-trivial:
\be
\delta\,\Gamma_\chi =   s_\chi\,\bigl(\delta\chi[\Phi]\bigr)
\ee
This is so  thanks to the $\chi$-dependence of the gauge-fixed BRST operator $s_\chi$.

Let us also observe that when $\chi[\Phi]$ does not depend on the reparametrization ghost $\xi^\mu$ one has
\be
s_\chi\, \chi[\Phi] = \bigl(S_\chi+ {\cal L}_\xi\bigr)\, \chi[\Phi]
\ee
where $S_\chi$ is the gauge-fixed $S$. Therefore, by taking $\chi[\Phi]$ 
invariant under simultaneous reparametrizations of both fields and backgrounds, one obtains
the gauge-fixed BV action
\be
\Gamma_\chi = \Gamma_{CS}[A] + S_\chi\,\bigl(\chi[\Phi]\bigr)-\frac{1}{ 2} \int_{M_3}{\rm Tr}\,\Bigl(i_\gamma\Bigl(\frac{\delta\,\chi}{ \delta\,A}\Big)\, \frac{\delta\,\chi}{ \delta\,A}\Bigr)
\label{gfactionbis}
\ee 
which is independent of the reparametrization ghost $\xi^\mu$.  Hence 
\be
S_\chi\, \Gamma_\chi =\bigl(S_0\, +G_\gamma\bigr)\, \Gamma_\chi =0
\label{Sinvariance}
\ee

\subsection{A digression: Chern-Simons global vector supersymmetry}
\label{globalvectorsusy}

The somewhat surprising outcome of our analysis so far is that
to put CS theory consistently on a curved manifold we must couple it not only to
the metric $g_{\mu\nu}$ and to its BRST partner $\psi_{\mu\nu}$ 
but also to the commuting vector field $\gamma^\mu$.

Coupling to classical background is one way to study the
(quantum) properties of the corresponding conserved currents
in {\it flat} space. In this subsection we make a digression from the main line of the paper
and pause to understand the necessity of the source $\gamma^\mu$
from the point of view of the global symmetries of CS
theory in flat space.

The background metric $g_{\mu\nu}$ is of course associated to
the conserved symmetric stress-energy tensor $T_{\mu\nu}$. Classically, the 
topological nature of the theory is expressed  by the relation
\be
T_{\mu\nu} = S_0\,S_{\mu\nu}
\ee
where $S_{\mu\nu}$ is associated to the source $\psi^{\mu\nu}$ and is obtained from the gauge-fermion functional
\be
S_{\mu\nu} = \frac{\delta\chi}{\delta\,g^{\mu
\nu}}
\ee
We would like to understand if  $S_{\mu\nu}$  correspond or not to 
symmetries of the theory in flat space and to which  current do 
the sources  $\gamma^\mu$ correspond to.  To answer these questions
let us look at the  $\gamma$-dependent part of the gauge-fixed action 
\bea
&&G_\gamma\, \bigl(\chi[\Phi]\bigr)-\frac{1}{ 2}\int_{M_3}{\rm Tr}\,\Bigl(i_\gamma\Bigl(\frac{\delta\,\chi}{ \delta\,A}\Big)\, \frac{\delta\,\chi}{ \delta\,A}\Bigr)=\nonumber\\
&&\qquad= \int_{M_3}{\rm Tr}\,\Big[\frac{\delta\chi}{ \delta c}\, i_\gamma (A) + \frac{\delta\chi}{ \delta \Lambda}\, {\cal L}_\gamma\, b-
\frac{1}{ 2}\,\epsilon^{\mu\nu\rho}\, \gamma^\sigma\,g_{\rho\sigma}\frac{\delta\chi}{\delta A_\mu}\frac{\delta\chi}{\delta A_\nu} \Bigr]
\label{gammaterms}
\eea 
Consider for concreteness Landau's gauge 
\be
\chi[\Phi] = \,\int_{M_3}{\rm Tr}\, \bigl[b\, \bar{D}^\dagger\,A\bigr] =\,\int_{M_3} \sqrt{g}\,b^{(0)}\, \bar{D}^\mu\,A_\mu\, d^3 x
\ee
where $\bar{D}^\dagger$ is the exterior differential which is Hodge-dual to
\be
\bar{D}=dx^\mu\, \bar{D}_\mu
\ee
$\bar{D}_\mu$ is the derivate covariant with respect to the Levi-Civita connection built with $g_{\mu\nu}$ and   $b^{(0)}$ is the 0-form
dual to $b$.
For such choice one has
\be
\frac{\delta\chi}{\delta c}=\frac{\delta\chi}{\delta \Lambda}=0
\label{landautypegauges}
\ee
Thus, the $\gamma$-dependent part of the gauge-fixed action  in Landau's gauge reduces to
\be
-\int_{M_3}
\frac{1}{ 2}\,\epsilon^{\mu\nu\rho}\, \gamma^\sigma\,g_{\rho\sigma}\,{\rm Tr}\,\frac{\delta\chi}{\delta A_\mu}\frac{\delta\chi}{\delta A_\nu}=-\int_{M_3}
\frac{1}{ 2}\,\epsilon^{\mu\nu\rho}\, \gamma^\sigma\,g_{\rho\sigma}\,{\rm Tr}\,\partial_\mu b^{(0)}\,\partial_\nu\, b^{(0)}
\label{landautypegaugesbis}
\ee
From this we see that  the gauge-fixed action in {\it
flat space} becomes independent of $\gamma^\mu$  when  $\gamma^\mu$ is  {\it constant}. In this limit, the BRST operator
\be
G_\gamma \to \gamma^\mu\, G_\mu
\ee
turns into a {\it global} vector supersymmetry which acts on the fields as follows
\bea
&&\gamma^\mu\,G_\mu\, c = -i_\gamma(A)\qquad \gamma^\mu\,G_\mu\, A =  - i_\gamma ( \bar{D}^\dagger b)
\label{gfbrstlandau}
\eea

This is the vector global supersymmetry of CS gauge-fixed action in Landau's gauge which was first discovered in \cite{Delduc:1989ft} and then extensively
studied,
for example, in \cite{Birmingham:1988ap,Delduc:1990je,Lucchesi:1992gp}. 
We see that, in a sense, such a global symmetry is the remnant, in flat space, of the equivariant
part of the topological BRST symmetry on curved manifolds: it is  associated to the super-ghost $\gamma^\mu$, which is required to render equivariant the coupling
of CS to topological gravity. Our discussion characterizes the most general gauges which enjoy global vector supersymmetry: these are the gauges for which the
$\gamma$-dependent terms of the action, given by Eq. (\ref{gammaterms}), vanish in flat space. 
Another example, beyond the Landau's gauge, is the axial gauge \cite{Brandhuber:1992ur} for
which, indeed, both (\ref{landautypegauges}) and (\ref{landautypegaugesbis}) hold
\footnote{For axial gauge (\ref{landautypegauges}) and (\ref{landautypegaugesbis}) would be valid
even in curved space, but the definition of the axial gauge requires the existence of a Killing vector, which restricts the possible background
metrics to the one with continuous isometries. In this case too therefore the vector supersymmetry is  a global one.}.

One might therefore think that $S_{\mu\nu}$ are the conserved  super-currents associated to the charges $G_\mu$. But this is not quite true, as we will see
momentarily. One can verify that the conserved super-currents $\tilde{S}_{\mu\nu}$ associated to flat space symmetries $G_\mu$ are not symmetric in the indices $\mu$ and $\nu$:
\bea
&&\tilde{S}_{\mu\nu} =S_{\mu\nu}+ \Delta_{\mu\nu} \qquad \partial^\mu\,\tilde{S}_{\mu\nu} =0\nn\\
&&S_{\mu\nu}=S_{\nu\mu}\qquad\qquad\quad  \Delta_{\mu\nu}= -\Delta_{\nu\mu}\eea
where
\bea
&&\Delta_{\mu\nu} = S_0 \, J_{\mu\nu}\qquad J_{\mu\nu} = \frac{1}{2} \epsilon_{\mu\nu\rho}\, {\rm Tr}\, D^\rho\,b^{(0)}\, b^{(0)}
\eea
Neither $S_{\mu\nu}$ nor $\Delta_{\mu\nu}$  is conserved. This is
consistent with conservation of the stress-energy tensor $T_{\mu\nu}$ since
the latter implies conservation of $S_{\mu\nu}$ only up to $S_0$-trivial terms:
\bea  &&T_{\mu\nu} = S_0\, \tilde{S}_{\mu\nu} =S_0\, S_{\mu\nu} \nn\\
&& 0=\partial^\mu\,  T_{\mu\nu}=
S_0\,\partial^\mu\, \tilde{S}_{\mu\nu} =S_0\, \partial^\mu\,S_{\mu\nu} \eea

Since the truly conserved super-currents $\tilde{S}_{\mu\nu}$ are not symmetric,
it is not possible to turn the global symmetries  of the theory 
into local ones, by only using sources $h^{\mu\nu}= g^{\mu\nu}-\delta^{\mu\nu}$ and $\psi^{\mu\nu}$ symmetric in the indices $\mu$ and $\nu$. One needs to add one more source, anti-symmetric in the indices $\mu$ and $\nu$
\bea && h^{\mu\nu}\, T_{\mu\nu} + \psi^{\mu\nu}\, S_{\mu\nu} +\phi^{\mu\nu}\, S_0\,J_{\mu\nu} +\cdots\nn\\
&&\delta h^{\mu\nu}= D^{(\mu}\,\xi^{\nu)}\qquad  \delta \psi^{\mu\nu}= D^{(\mu}\,\gamma^{\nu)}\quad \delta \phi^{\mu\nu}= D^{[\mu}\,\gamma^{\nu]} 
\eea
Working out the Noether procedure in a way compatible with $S_0$-invariance one is led  to the  coupling 
\be D^{[\mu}\,\gamma^{\nu]}\, J_{\mu\nu} = \frac{1}{2}
\bigl(D^\mu\,\gamma^\rho\bigr)\,\epsilon_{\mu\nu\rho}\, {\rm Tr}\, b^{(0)}\,D^\nu\, b^{(0)}
\nn\ee
that was found with the BV method.

In conclusion, from the point of view of the conserved currents of the theory in flat space the coupling to topological gravity encodes the relations
\bea 
&&\partial^\mu\, T_{\mu\nu}=0\qquad T_{\mu\nu}= S_0\, S_{\mu\nu}\qquad\partial ^\mu\, S_{\mu\nu} = - S_0\, \partial^\mu\, J_{\mu\nu}
\nn\eea

\sectiono{A Deformation of Chern-Simons Theory}
\label{higherghostCS}

Let us collect the fields and anti-fields of the gauge sector into 
a generalized form with values in the Lie algebra of $G$:
\be
{\cal A} \equiv c+ A+ A^*+ c^*
\ee
The usefulness of introducing such a field with indefinite form degree in Chern-Simons
theory has been pointed out in \cite{Axelrod:1993wr} and 
\cite{Alexandrov:1995kv}.

${\cal A}$ is {\it odd} since has total fermionic degree (form+ghost number) $f=1$. 
As reviewed in the Appendix \ref{equivariantcohomology} the equivariant 
cohomology of $s$ modulo $d$ is equivalent to the cohomology of
the coboundary operator $\delta$
\be
\delta \equiv S+i_\gamma- d
\label{equivariantdeltabis}
\ee
on the space of generalized forms which do not contain the ghost 
field $\xi^\mu$. Observe that Eqs. (\ref{ggammamatter}) are equivalent to 
\be
\bigl(G_\gamma\,+ i_\gamma \bigr)\,{\cal A} = 0
\label{chiralA}
\ee

Therefore the  action of $\delta$ on it 
coincides with that of  the ``naive'' $\delta_0\equiv  S_0 - d$:
\be
\delta\,{\cal A} = \delta_0\, {\cal A}
\ee
The BRST transformations (\ref{brsmatterequivariant}) rewrite as follows
\be
\delta\, {\cal A} =  \delta_0\, {\cal A}= {\cal A}^2
\label{brstageneralized}
\ee

It is well known that elements of the local $\delta_0$-cohomology 
made of ${\cal A}$ are built with the help  of $G$-invariant
anti-symmetric polynomials in the Lie algebra of $G$
\cite{DuboisViolette:1985jb}. 
If $\tau_{a_1,\ldots a_m}$  is a $G$-invariant
completely anti-symmetric tensor with $m$ indices $a_i$ 
running in the adjoint of $G$, the generalized form
\be
\langle {\cal A}^m\rangle \equiv \tau_{a_1,\ldots a_m}\, {\cal A}^{a_1}\ldots
{\cal A}^{a_m}
\label{minvariants}
\ee
corresponds to a non-trivial class of $\delta_0$. 
Thanks to the ``chirality'' property (\ref{chiralA}) of ${\cal A}$,
this is also a  class  of the  equivariant $\delta$. 

The smallest value of $m$ is $m=3$, for which there is a unique
choice $\tau_{abc}= f_{abc}$, the structure constants of the Lie algebra of $G$.  One can write the corresponding observable
\be
\Omega_3\equiv \frac{1}{ 3}{\rm Tr}\, {\cal A}^3\equiv \Omega_3^{(0)}+\Omega_2^{(1)}+\Omega_1^{(2)}+\Omega_0^{(3)}
\ee
in terms of traces over some irreducible representation $R$, the dependence
on the representation being an overall multiplicative factor. 
The forms
\bea
&& \Omega^{(0)}_3 = \frac{1}{ 3}\, {\rm Tr} \, c^3\nonumber\\
&&\Omega^{(1)}_2= {\rm Tr}\, A\, c^2\nonumber\\
&&\Omega^{(2)}_1 = {\rm Tr}\bigl[A^*\, c^2 +A^2\, c\,\bigr]\nonumber\\
&&\Omega^{(3)}_0 = \frac{1}{3}\, {\rm Tr}\bigl[ A^3 + 3\,c^*\, c^2 + 3\, \{A,c\} \,A^*\bigr]
\eea
satisfy the descent equations (\ref{sgammadescent}). In particular the integral
of the 3-form  has ghost number 0 and is  both $S$ and $S_0$ invariant
\be
S \int_{M_3} \Omega^{(3)}_0 = S_0 \int_{M_3} \Omega^{(3)}_0 = G_\gamma\int_{M_3} \Omega^{(3)}_0 =0
\ee
This cocycle is related to the CS classical action by the equation
\be
\omega_{CS}=-\frac{1}{ 3!}{\rm Tr} {\cal A}^3   + \delta_0 \alpha
\ee
where $\omega_{CS}$ is the $\delta_0$-cocycle
\bea
&&\omega_{CS} \equiv \omega_{CS}^{(0)}+\omega_{CS}^{(1)}+\omega_{CS}^{(2)}+
\omega_{CS}^{(3)}
\eea
whose 3-form component is the CS form 
\bea
&&\omega^{(3)}_{CS} ={\rm Tr}\,\bigl[ \frac{1}{2}\,A\,dA+\frac{1}{ 3} A^3\bigr]=-\frac{1}{ 2!}\Bigl[\Omega^{(3)}_0+d\,{\rm Tr}\, A^*\, c - S_0\,{\rm Tr} \,\bigl[A^*\, A + c^*\,c\bigr]\Bigr]\nn\\
&&\omega^{(2)}_{CS} = 
\frac{1}{ 2!}\, dA\,c=-\frac{1}{ 2!}\,\Bigl[\Omega^{(2)}_1-S_0\, {\rm Tr}\, A^*\, c\Bigr]\nonumber\\
&&\omega^{(1)}_{CS} = -\frac{1}{ 2!}\,\Omega^{(1)}_2 \nonumber\\
&&\omega^{(0)}_{CS}= -\frac{1}{ 2!}\,\Omega^{(0)}_3 
\eea
and $\alpha$ is a generalized form of total ghost number $f=2$:
\bea
&&\alpha = \frac{1}{ 2!}\,{\rm Tr}\,\bigl[ A^*\, c+ A^*\, A + c^*\,c\bigr]
\eea
Thus $\Omega_0^{(3)}$ and the CS classical form $\omega^{(3)}_{CS}$ are equivalent (up to a multiplicative factor) 
in the ``naive'' $S_0$ cohomology (modulo d). However, $\alpha$ 
does not satisfy the ``chirality'' condition (\ref{chiralA}). Therefore $\omega_{CS}^{(3)}$ is  not an equivariant cocycle, and it is 
{\it not} equivalent to $\Omega_0^{(3)}$ in the equivariant 
$s$-cohomology (modulo d).
As we verified in  Section 3, to construct an equivariant action 
 starting from the classical
CS form, one must introduce  explicit couplings 
to the topological gravity fields.

Since $\Omega_0^{(3)}$ is $S$-invariant modulo $d$ we can add it to the 
BV action
\be
\tilde{\Gamma}_{BV} (t) = \Gamma_{BV} +t \int_{M_3} \Omega^{(3)}_0
\label{deformedactiont3}
\ee
The new BV action generates {\it deformed} BRST transformation rules.
It is straightforward to verify that the new action is 
just the rescaled original BV action
\be
\tilde{\Gamma}_{BV}(t) =  \frac{1}{(1+t\bigr)^2}\, \Gamma_{BV}({\mathcal A})\Big|_{{\cal A}\to 
\bigl(1+ t\bigr)\,{\mathcal A} }
\ee
In this sense, the effect of adding this observable to the BV action is just 
to renormalize both the coupling constant and the fields. 

Following the suggestion of \cite{Witten:1992fb} we will now consider
cocycles with $m>3$. The number and structure of
the invariant tensors $\tau_{a_1,\ldots,a_m}$ with $m>3$ 
depend on the gauge group $G$. In the rest of this Section 
we will restrict  our discussion to $G=SU(N)$, both for concreteness and because we have in mind an application to topological D-branes.

A special class of $SU(N)$-invariant tensors
are anti-sym\-me\-trized single traces of an {\it odd} number of $SU(N)$ generators 
in some irreducible representation $R$
\be
\tau^{(R)}_{a_1,\ldots a_m} = {\rm Tr}_R T^{[a_1}\cdots T^{a_m]}
\label{antisymsingletrace}
\ee
If  $m$ takes one of the following $N-1$ values
\be
m= 3, 5, 7,\ldots, 2\,N-1
\label{mrange}
\ee
the dependence of $\tau^{(R)}_{a_1,\ldots a_m}$ on the representation $R$ 
is a multiplicative  --- possibly vanishing --- normalization factor\footnote{The factor  $d_m(R)$ vanishes for $R=\bar{R}$ and $m=5,9,11,\ldots$}:
\be
\tau^{(R)}_{a_1,\ldots a_m} = d_m(R)\, \tau_{a_1,\ldots a_m}
\ee
with $\tau_{a_1,\ldots a_m}$ independent of $R$. 

All the other $SU(N)$ anti-symmetric tensors are multi-traces.
They are obtained by multiplying and anti-symmetrizing single
trace invariants  with $m$ in the range (\ref{mrange}). In the following
we will therefore restrict ourselves to such ``primitive'' single trace
invariants.

The observables corresponding to the single trace invariants are
\be
\Omega_m = \frac{1}{ m}\,{\rm Tr}\, {\cal A}^m \equiv \Omega_m^{(0)}+\Omega_{m-1}^{(1)}+
\Omega_{m-2}^{(2)}+\Omega_{m-3}^{(3)}
\ee
where
\bea
&& \!\!\!\Omega^{(0)}_m = \frac{1}{ m}\, {\rm Tr} \, c^m\nonumber\\
&&\!\!\!\Omega^{(1)}_{m-1}= {\rm Tr}\, A\, c^{m-1}\nonumber\\
&&\!\!\!\Omega^{(2)}_{m-2} =  {\rm Tr}\,\Bigl[
A^*\, c^{m-1} +\frac{1}{ 2}\bigl(
A^2\, c^{m-2}+A\, c\, A \, c^{m-3}+\cdots +A\, c^{m-2}\, A\bigr)\Bigr]\nonumber\\
&&\!\!\!\Omega^{(3)}_{m-3} =  {\rm Tr}\,\Bigl[ c^*\,c^{m-1} + \frac{1}{ 2}\,A^*\,\bigl(A\, c^{m-2}+ c\, A\,c^{m-3}+\cdots  +c^{m-2}\, A\bigr)+\nonumber\\
&&\qquad \qquad +\sum_{i,j \le {m-3}}A\, c^{m-i}\, A\, c^{j}\, A\, c^{i-j-3} \Bigr]
\eea
The 3-forms $\Omega^{(3)}_{m-3}$ have  ghost number $m-3$. By adding them 
to the original BV action one obtains the generic single trace deformation 
of $SU(N)$ CS theory
\be
\tilde{\Gamma}_{BV}(t_i) =   \Gamma_{BV}+ \sum_{i=1}^{N-1}
t_i\, \int_{M_3}\Omega^{(3)}_{2\,i}
\label{deformedBV}
\ee
The new BV action generates deformed nilpotent BRST transformation 
rules which can be recast as follows
\bea
&&\delta_t\, {\cal A} = \{\tilde{\Gamma}_{BV}(t_i),{\cal A}\}= 
{\cal A}^2 + \sum_{i=1}^{N-1}\, t_i\, {\cal A}^{2\,i+2}\qquad\nn\\
&& \delta_t^2=0
\eea
where
\be
\bigl({\cal A}^{m-1}\bigr)^a \equiv g^{aa_1}\, \tau_{a_1 a_2\ldots a_m}\, {\cal A}^{a_2} \cdots {\cal A}^{a_m}
\ee 
and $g^{ab}$ is the invariant Killing metric on the Lie algebra of $G$.

\sectiono{Effective Action}
\label{effectiveaction}

Quantum averages of matter observables are formally obtained by integrating
the gauge-fixed BV path integral over the fields of the gauge sector
\be
{\rm e}^{i\,\frac{k}{2\,\pi}\,F[g_{\mu\nu},\psi_{\mu\nu},\gamma^\mu; t_i]}= \int \bigl[dA\,dc\,db\,d\,\Lambda]\, {\rm e}^{i\, \frac{k}{2\,\pi}\, \tilde{\Gamma}_{BV}(t_i)
\big|_{\Phi^*=\frac{\delta\,\chi}{ \delta\,\Phi}}}
\ee
The  classical Ward identities for the effective action 
 $F[g,\psi,\gamma;t]$ read as follows
\be
S\,F[g_{\mu\nu},\psi_{\mu\nu},\gamma^\mu; t_i]= \int_{M_3}\Bigl[ \frac{\delta\, F}{ \delta g_{\mu\nu}}\, \psi_{\mu\nu} - \frac{\delta\, F}{ \delta \psi_{\mu\nu}}\, {\cal L}_\gamma\,g_{\mu\nu}\Bigr]=0
\label{widentityclassical}
\ee
Let us consider $G=SU(N)$. 
Conservation of total ghost number implies
\be
F[g_{\mu\nu},\psi_{\mu\nu},\gamma^\mu; t_i] = \sum_{n=0}^\infty F_{2\,n}[g_{\mu\nu},\psi_{\mu\nu},\gamma^\mu;t_i]
\label{effectivegravity}
\ee
where $F_{2\,n}[g_{\mu\nu},\psi_{\mu\nu},\gamma^\mu;t_i]$ is a functional of the
topological gravity fields of ghost number $2\,n$ which is polynomial in the couplings
$t_i$:
\bea
&& F_{2\,n}[g_{\mu\nu},\psi_{\mu\nu},\gamma^\mu;t_i]=\!\!\!\!\!\!\sum_{\sum_\alpha i_\alpha= n}\!\!\!\!{\cal F}_{i_1; i_2;\ldots ;i_\alpha;\ldots}[g_{\mu\nu},\psi_{\mu\nu},\gamma^\mu]\, 
t_{i_1}\cdots t_{i_\alpha}\cdots
\label{effectiveactions}
\eea
In other words the observables that contribute to the effective action of fixed ghost number $2\,n$ are those which satisfy the selection rule
\be
\sum_\alpha i_\alpha = n\qquad   1\le i_\alpha \le N-1
\label{ghostselectionrule}
\ee
Therefore for finite $N$ there is only a finite number of terms in the sums (\ref{effectivegravity}) and (\ref{effectiveactions}).

Let us recall \cite{Ouvry:1988mm,Becchi:1995ik} the geometrical interpretation of the classical Ward identity (\ref{widentityclassical}). 
Let ${Met}_3$ be the space of 3-dimensional metrics on $M_3$,  $Diff(M_3)$, the diffeomorphisms of $M_3$ and 
\be
{\cal M}={Met}_3/Diff(M_3)
\ee
the --- possibly infinite-dimensional --- associated orbit space. We will refer to it as the  moduli space of 3-dimensional metrics. 

Let $m\equiv \{m^a\}$ be local coordinates on ${\cal M}$ and  $\bar{g}_{\mu\nu}(x; m)$ 
a local section of the bundle ${\cal B}$ whose total space is ${Met}_3$ and whose base is ${\cal M}$.  Let 
\be
d_m \equiv d\, m^a\,\frac{\partial}{\partial m^a}
\ee
the exterior derivative on ${\cal M}$. When acting on  
$\bar{g}_{\mu\nu}(x; m)$, $d_m$ does not in general produce 
a tensor covariant under $m$-dependent diffeomorphisms of $M_3$. One has
\be
d_m \bar{g}_{\mu\nu}(x,m) = \bar{\psi}_{\mu\nu}(x;m) - {\cal L}_{\bar{\xi}}\, 
\bar{g}_{\mu\nu}(x;m)
\label{dmmodulione}
\ee
where $\bar{\psi}_{\mu\nu}(x,m)$ is covariant under $m$-dependent  $M_3$-diffeomorphisms --- i.e. it is
a local section of the cotangent bundle $T^*\,{\cal M}$ --- and ${\bar{\xi}}^\mu(x;m)$
is an $m$-dependent vector field on $M_3$ with values in the local 1-forms on 
${\cal M}$. ${\bar\xi}^\mu(x;m)$ defines a local connection on the bundle ${\cal B}$. 

Nilpotency of  $d_m$ implies
\be
d_m \bar{\psi}_{\mu\nu}(x,m) = {\cal L}_{\bar{\gamma}}\,\bar{g}_{\mu\nu}(x;m) - 
{\cal L}_{\bar{\xi}}\, \bar{\psi}_{\mu\nu}(x;m)
\label{dmmodulitwo}
\ee
where $\bar{\gamma}^\mu(x;m)$ is a $m$-dependent vector field on $M_3$ with values in the 2-forms on ${\cal M}$.  $\bar{\gamma}^\mu(x;m)$ is related to the curvature 2-form of the connection $\bar{\xi}^\mu(x;m)$ 
\be
d_m \bar{\xi}^\mu(x;m)+ \frac{1}{ 2}{\cal L}_{\bar{\xi}}\, \bar{\xi}^{\mu}(x;m)=\bar{\gamma}^{\mu}(x;m) 
\label{dmmodulithree}
\ee
Thus the action of $d_m$ on 
the local sections $\bar{g}_{\mu\nu}(x;m)$, $\bar{\psi}_{\mu\nu}(x;m)$ and  $\bar{\gamma}^\mu(x;m)$ is identical to the action (\ref{brsgravityequivariant}) of the equivariant BRST operator $s$ on the topological
gravity fields.

In conclusion, the classical  Ward identity (\ref{widentityclassical})
states that the effective action $F_{2n}$, when evaluated on the backgrounds 
$\bar{g}_{\mu\nu}(x;m),\bar{\psi}_{\mu\nu}(x;m)$ and $\bar{\gamma}^\mu(x;m)$, gives rise to a {\it local  closed} $2n$-form 
${\bar F}_{2n}$ on  moduli space ${\cal M}$. 

The equivariance condition ensures that the 
local form ${\bar F}_{2n}$ is actually {\it globally} defined on  ${\cal M}$. In fact, equivariance means that 
${\bar F}_{2n}$ is independent of the choice of the connection $\bar{\xi}^{\mu}(x;m)$ which had to be 
introduced in order to define $\bar{\psi}_{\mu\nu}(x;m)$ and $\bar{\gamma}^\mu(x;m)$. 
Since $\bar{\psi}_{\mu\nu}(x;m)$ and $\bar{\gamma}^\mu(x;m)$ transform covariantly under $m$-dependent diffeomorphisms --- unlike  $\bar{\xi}^{\mu}(x;m)$ ---  the  local form ${\bar F}_{2n}$ extends to a globally defined form on  ${\cal M}$. 

As explained in \cite{Witten:1992fb}, it makes sense, in this situation, to integrate the closed, globally defined form  ${\bar F}_{2n}$ on $2\,n$-dimensional cycles $C_{2\,n}$ of moduli space ${\cal M}$ 
\be
I_{2n}(t_i)=\int_{C_{2\,n}}\!\!\!\!\!\! F_{2n} [\bar{g}_{\mu\nu}(x;m),\bar{\psi}_{\mu\nu}(x;m), \bar{\gamma}^\mu(x;m);t_i]
\label{topinvariants}
\ee
Were the classical Ward identity satisfied, the $I_{2n}(t_i)$'s would be 
topological invariants of the 3-dimensional manifold  $M_3$ depending on the homology class of $C_{2\,n}$. In the next Section we will consider the possible quantum
anomalies that can appear on the right-hand side of the classical Ward identity
(\ref{widentityclassical}).

\sectiono{Topological Anomalies}
\label{topanomalies}

Anomalous Ward identities for the $G=SU(N)$ theory
\be
S\, F_{2n}[g,\psi,\gamma] = \int_{M_3}\,A^{(3)}_{2\,n+1}[g, \psi, \gamma]
\ee
involve local functionals of the topological gravity fields of ghost number
$2\,n+1$ which are $S$-closed modulo $d$:
\be
S\, A^{(3)}_{2\,n+1} = d\, A^{(2)}_{2\,n+2}
\ee
$A^{(2)}_{2\,n+2}$ is a 2-form of ghost number $2\,n+2$. 
This implies that $A^{(3)}_{2\,n+1}$ is the 3-form component of a generalized
anomaly form
\be
{\cal A}_{2\,n+4} = A^{(3)}_{2\,n+1}+ A^{(2)}_{2\,n+2}+A^{(1)}_{2\,n+3}+A^{(0)}_{2\,n+4}
\ee
which is $\delta$-closed
\be
\delta\, {\cal A}_{2\,n+4}=0
\label{deltacocycles}
\ee
Local solutions of (\ref{deltacocycles}) modulo $\delta$ are local observables of
3-dimensional topological gravity. Construction of topological gravity 
observables 
in various dimensions has been discussed by several authors, starting from
\cite{Myers:1989dn,Myers:1990zi}. Although the formalism developed
in \cite{Myers:1989dn} applies to any dimension,  that work and most of the 
subsequent ones focused mainly on either  2 or (less so) 4 dimensions. 
In particular, as far as we know, no explicit formulae have been exhibited in dimension 3.  We think therefore it is appropriate to present here a review of the relevant results\footnote{We believe that our discussion is slightly more general than the original one \cite{Myers:1989dn}, since we explicitly show in the
Appendix \ref{filtration}
that the trace classes we consider do not depend on the choice of the particular curvature two-form, be it with or without ``torsion''.}

To find local solutions of Eq. (\ref{deltacocycles}) we start from the matrix-valued curvature 2-form 
\be
(R^{(2)})^{\mu}_{\;\nu} = \frac{1}{ 2} (R_{\alpha\beta})^{\mu}_{\;\nu}\, dx^\alpha\,dx^\beta=
(d\,\Gamma+ \Gamma^2)^{\mu}_{\;\nu}
\ee
where $\Gamma$ is the matrix-valued 1-form
\be
(\Gamma)^{\mu}_{\;\nu}= \Gamma^{\mu}_{\alpha\nu}\,dx^\alpha
\ee
From 
\be
{\cal L}_\gamma \Gamma = D\, R^{(0)} - i_\gamma R^{(2)}
\ee 
one derives the following {\it matrix-valued} descent equations:
\bea
&&S\, R^{(2)} = D\, R^{(1)}\nonumber\\
&&S\, R^{(1)} = D\, R^{(0)} - i_\gamma R^{(2)}\nonumber\\
&& S\, R^{(0)} = -i_\gamma R^{(1)}
\label{matrixvalueddescent}
\eea
where we introduced the matrix-valued 1 and 0-forms
\bea
&& (R^{(1)})^{\mu}_{\;\nu} \equiv S\, \Gamma^{\mu}_{\;\nu}= \frac{1}{ 2}\bigl[ D\, \psi^{\mu}_{\;\nu}+ D_\nu\, \psi^{\mu}_{\;\alpha}\,dx^\alpha- D^\mu\,\psi_{\alpha\nu}\,dx^\alpha\Bigr] \nonumber\\
&& (R^{(0)})^{\mu}_{\;\nu}\equiv D_\nu\,\gamma^\mu\eea
Let us define the  matrix-valued generalized form of total fermionic degree 2
\be
{\cal R} \equiv R^{(2)}+R^{(1)}+R^{(0)}
\ee
and the differential 
\be
\delta \equiv S +i_\gamma - D
\ee
acting on matrix-valued generalized forms  $X$.  It easy to see that
\be
\delta^2 X = \bigl[{\cal R} ,X\bigr]
\ee
Hence $\delta$ is nilpotent when acting on reparametrization scalars. 
Moreover one has
\be
\delta\,  {\cal R}=0
\ee
which is equivalent to Eqs. (\ref{matrixvalueddescent}). The invariants constructed with
${\mathcal R}$
\be
{\cal A}_{2\,k}= {\rm tr}\, {\cal R}^{k}\qquad k=1,2,\ldots
\label{topgrcocycles}
\ee
where the trace is taken on the matrix Lorentz indices, are therefore $\delta$-cocycles of total ghost number $2\,k$:
\be
\delta\, {\cal A}_{2\,k}=0
\label{topgrcocyclesbis}
\ee
Let us  discuss if they are trivial or not. To this end, consider the decomposition of  the matrix ${\cal R}$
\be
{\cal R} \equiv \tilde{\mathcal R} + \tilde{\mathcal F}
\label{decompositioncurv}
\ee
into its symmetric and antisymmetric parts,  with respect to $g^{\mu\nu}$:
\bea
&&\bigl(\tilde{\mathcal R})^\mu_{\;\nu}\,g^{\nu\lambda}  =  -g^{\mu\nu}
\,\bigl(\tilde{\mathcal R})^\lambda_{\;\nu}\nn\\
&&\bigl(\tilde{\mathcal F})^\mu_{\;\nu}\,g^{\nu\lambda}  =  g^{\mu\nu}\,
\bigl(\tilde{\mathcal F})^\lambda_{\;\nu}
\eea
Explicitly:
\bea
&& (\tilde{R}^{(2)})^{\mu}_{\;\nu} =(R^{(2)})^{\mu}_{\;\nu} \nn\\
&& (\tilde{R}^{(1)})^{\mu}_{\;\nu} = \nonumber \frac{1}{ 2}\bigl[ D_\nu\, \psi^{\mu}_{\;\alpha}\,dx^\alpha- D^\mu\,\psi_{\alpha\nu}\,dx^\alpha\Bigr] \\
&& (\tilde{R}^{(0)})^{\mu}_{\;\nu}=\frac{1}{2}\,\bigl(D_\nu\,\gamma^\mu-D^\mu\,\gamma_\nu\bigr)
\eea
and
\bea
&& (\tilde{\mathcal F}^{(2)})^{\mu}_{\;\nu} =0 \nn\\
&& (\tilde{\mathcal F}^{(1)})^{\mu}_{\;\nu} =  \frac{1}{ 2}\,D\, \psi^{\mu}_{\;\nu} \nonumber\\
&& (\tilde{\mathcal F}^{(0)})^{\mu}_{\;\nu}=\frac{1}{2}\,\bigl(D_\nu\,\gamma^\mu+D^\mu\,\gamma_\nu\bigr)
\eea
The important observation is 
\bea
&&\tilde{\mathcal F}^\mu_\nu = -\frac{1}{2}\,g^{\mu\lambda}\, \delta \psi_{\nu\lambda}=-\delta {\mathcal C}^\mu_\nu + \frac{1}{2} \delta\bigl(g^{\mu\lambda}\bigr)\,\psi_{\nu\lambda}=\nn\\
&&\qquad =-\delta\,{\mathcal C}^\mu_\nu - \frac{1}{2}\,\psi^{\mu\lambda}\, \psi_{\nu\lambda}=-\bigl(\delta\,{\mathcal C} +{\mathcal C}^2\bigr)^\mu_\nu - \frac{1}{4}\,\psi^{\mu\lambda}\, \psi_{\nu\lambda}
\eea
where ${\mathcal C}$ is the matrix of fermionic number +1:
\be
{\mathcal C}^\mu_\nu \equiv \frac{1}{2}\, \psi^\mu_\nu
\ee
The decomposition (\ref{decompositioncurv}) rewrites therefore as follows
\be
{\cal R} = \hat{\mathcal R} - {\mathcal F}
\label{decompositioncurvbis}
\ee
where
\bea
&&\hat{\mathcal R}  \equiv \tilde{\mathcal R} -\frac{1}{4}\,\psi^{\mu\lambda}\, \psi_{\nu\lambda}\qquad  {\mathcal F}\equiv \delta\,{\mathcal C} +{\mathcal C}^2
\eea
Since $\psi^\mu_\nu$ is anti-commuting, $\hat{\mathcal R}$ is anti-symmetric with respect to $g^{\mu\nu}$:
\bea
&&\bigl(\hat{\mathcal R})^\mu_{\;\nu}\,g^{\nu\lambda}  =  -g^{\mu\nu}\,
\bigl(\hat{\mathcal R})^\lambda_{\;\nu}
\eea
It should be noted, however, that, $ {\mathcal F}$ is not symmetric with respect to $g^{\mu\nu}$.

Since 
\be
\delta\, {\mathcal F}= \delta^2\, {\mathcal C} +\bigl[\delta\,{\mathcal C} ,{\mathcal C} \bigr]=\bigl[\hat{\mathcal R},{\mathcal C}\bigr]
\label{deltaF}
\ee
one has
\be
\delta\,\hat{\mathcal R}=- \bigl[{\mathcal C},\hat{\mathcal R}\bigr]
\label{deltarhat}
\ee
It is therefore convenient to introduce a new differential $\hat{\delta}$ whose action on generalized matrix-valued forms $X$ is
\be
\hat\delta\, X \equiv \delta\, X + \bigl[{\mathcal C}, X\bigr]
\label{deltahatX}
\ee
The curvature of $\hat\delta$ is $\hat{\mathcal R}$
\be
\hat{\delta}^2\, X= \bigl[\hat{\mathcal R}, X\bigr]
\ee
Eq. (\ref{deltarhat}) becomes the Bianchi identity for $\hat{\mathcal R}$:
\be
\hat{\delta}\,\hat{\mathcal R}=0
\ee
and Eq. (\ref{deltaF}) rewrites
\be
\hat{\delta}\,{\mathcal F}=\bigl[{\mathcal R},{\mathcal C}\bigr]
\ee
Hence the invariants constructed with $\hat{\mathcal R}$ 
\be
\hat{\mathcal A}_{2\,k}={\rm tr}\, \hat{\mathcal R}^{k}
\label{topgrcocycleshat}
\ee
are $\delta$-closed:
\be
\delta\, \hat{\mathcal A}_{2\,k}= 0
\label{topgrcocycleshatbis}
\ee
Comparison with (\ref{topgrcocycles}) gives
\be
\delta\bigl(\hat{\mathcal A}_{2\,k}-{\mathcal A}_{2\,k}\bigr) = 0 
\ee
One can check that the differences
\bea
&&{\mathcal B}_{2\,k} \equiv \hat{\mathcal A}_{2\,k}- {\mathcal A}_{2\,k} =  {\rm tr}{\mathcal F}^k +k\, {\rm tr}{\mathcal F}^{k-1}\, {\mathcal R}+\cdots
\eea
are {\it trivial} cocycles and thus
\be
 {\mathcal A}_{2\,k} \equiv \hat{\mathcal A}_{2\,k}\qquad {\rm in}\; \delta{\rm -cohomology}
\ee
We show this, for any space-time dimensions, in the Appendix
\ref{filtration}.  In the Appendix
\ref{filtration} we show also a more general proposition: The
non-vanishing cohomology of $\delta$ on the space of invariant
polynomials built with the matrices $\hat{\mathcal R}$, $g^{\mu\nu}$ and
${\mathcal F}$ is generated by the invariant polynomials built with of
$\hat{\mathcal R}$ and $g^{\mu\nu}$.  The trace classes $ {\mathcal A}_{2\,k}$ do not involve explicitly the metric, but only the
curvature forms.  Therefore these classes can be equivalently written in
terms of either the ${\mathcal R}$ or the $\hat{\mathcal R}$ curvature
form\footnote{The original works \cite{Myers:1989dn,Myers:1990zi} studying
topological gravity observables focused on even dimensions 2 and 4. In
even dimensions one can consider, beyond the trace classes, the Euler
class. This class involves the metric and therefore it is more
conveniently written in terms of $\hat{\mathcal R}$ and $g^{\mu\nu}$:
It necessarily contains ${\mathcal F}$ when expressed in terms of the
full curvature form ${\mathcal R}$. For this reason the authors of
 \cite{Myers:1989dn,Myers:1990zi}  chose a formalism in which 
$\hat{\mathcal R}$ and $g^{\mu\nu}$, but not 
${\mathcal R}$, appear.}.

Since $\hat{\mathcal R}$ is anti-symmetric with respect to
$g^{\mu\nu}$ the traces of the odd powers of $\hat{\mathcal R}$ vanish identically. In conclusion, in generic dimensions,  we end up with the 
following non-trivial equivariant cocycles 
\be
{\cal A}_{4\,p}\equiv\hat{\mathcal A}_{4\,p}\equiv{\rm tr}\, {\cal R}^{2\,p}\qquad p=1,2,\ldots
\label{topgrnontrivial}
\ee

Let us now specialize this discussion to the case at hand, i.e. to dimension 3. In such dimension  traces of powers of $\hat{\mathcal R}$ are all proportional to powers of 
${\rm tr}\,\hat{\mathcal R}^2$. Hence  
there exists one single non-trivial cocycle for any given total fermionic degree $4\, p$ 
\be
{\cal A}_{4\,p }\sim {\cal A}_4^p \equiv\bigl({\rm tr}\, {\cal R}^{2}\bigr)^p
\label{topgranomalies}
\ee
The corresponding CS topological anomaly has ghost 
number $4\,p-3$:
\be
S\, F_{4\,(p-1)}[g,\psi,\gamma] = c_{2\,(p-1)}(t) \int_{M_3} \bigl({\rm tr}\, {\cal R}^{2}\bigr)^p\qquad p=1,2,\ldots
\label{higherghostanomalies}
\ee
where the integration over $M_3$ selects the component of the 
anomaly of form degree 3 . For $p=1$ we obtain the well-known  {\it framing anomaly} of CS theory \cite{Witten:1988hf,BarNatan:1991rn}:
\be
A^{(3)}_{1} =   {\rm tr}\, R^{(2)} \, R^{(1)}=-\epsilon^{\mu\nu\rho}\,
R_\rho^\sigma\, D_\mu\,\psi_{\nu\sigma}\, d^3\,x
\ee
The anomalies with $p>1$ involve the super-ghost $\gamma^\mu$. They are relevant
for the higher-ghost deformations of  CS theory  discussed in Section \ref{higherghostCS}.

The cohomological analysis does not determine, of course, the coefficient of the possible
anomalies. It is known \cite{Witten:1988hf}  that the coefficient $c_0$ of the anomaly for the ghost number zero effective action $F_0$ does not vanish. For $G=SU(N)$  at 1-loop 
\be c^{1-loop}_0 =  \frac{1}{12}\, \frac{N^2-1}{k}\nn
\ee
where $\frac{1}{k}$ is the CS coupling constant. Witten \cite{Witten:1988hf} proposed an exact formula for $c_0$ motivated by the Hamiltonian solution of the theory in terms of CFT:
\be c_0 =  \frac{1}{12} \frac{N^2-1}{k +N}\nn\ee
This formula has been verified at 2-loops by an explicit perturbative computation \cite{Axelrod:1993wr}.

The coefficients $c_{2\,(p-1)}$ of the anomaly of $F_{4\,(p-1)}$ for $p>1$
are homogeneous polynomials of degree $2\,(p-1)$ in the $t_i$'s, with
$t_i$ of weight $i$:
\bea
&& c_2(t_1,t_2) = c^{(1)}_2\, t_2+ c^{(2)}_2\, t_1^2\nn\\
&& c_4(t_1,t_2, t_3,t_4) =c^{(1)}_4\, t_4+ c^{(2)}_4\, t_2^2+c^{(3)}_4\, t_2\, t_1^2+ c^{(4)}_4\, t_3\, t_1+ c^{(5)}_4\, t_1^4\nn\\
&&\cdots\qquad \cdots
\eea
Thus, the polynomials $I_{2\,n}(t_i)$  defined in Eq. (\ref{topinvariants}) are genuine 3-di\-men\-sio\-nal topological invariants for any $n$ {\it odd}.  Computation of the anomaly coefficients $c_{2\,(p-1)}(t)$ for $p >1$ appears to be an interesting open problem.  Since $c_{2\,(p-1)}(t)$  with $p>1$ receives contributions from more than one higher-ghost matter observable,  
$I_{2\,n}(t_i)$ are genuine topological invariants  for $n$ {\it even} as well when restricted on some suitable non-trivial sub-manifold of $t_i$-space. 

Let us conclude this Section with the following remark. The framing anomaly is locally
trivial:
\be
 A^{(3)}_1  = S\, \Omega^{(3)}_{0}(\Gamma) + d\,\Omega^{(2)}_{1}(\Gamma) 
\label{localtrivialitytwo} 
\ee
where
\be
\Omega^{(3)}_{0}(\Gamma)  \equiv \frac{1}{ 2} {\rm tr}\, \Gamma\, d\Gamma +\frac {1}{ 3}  {\rm tr}\,\Gamma^3
\ee
is the gravitational CS action and
\be
\Omega^{(2)}_{1}\equiv\frac{1}{ 2}{\rm tr}\,\Gamma\, R^{(1)}
\ee
This is best understood by considering the generalized form ${\cal A}_4$
in arbitrary dimension (i.e. different than 3). Its component with highest form degree 
is a 4-form
\be
\bigl({\cal A}_4\bigr)^{(4)}= A^{(4)}_0= \frac{1}{ 2}\, {\rm tr}\, (R^{(2)})^2
\ee
which is $d$-closed, by virtue of the Bianchi identity, and hence locally $d$-exact 
\be
A^{(4)}_0=d\, \Omega^{(3)}_{0}(\Gamma) 
\label{localtrivialityone}
\ee
Therefore
\bea
&& S\, \frac{1}{ 2} {\rm tr}\, (R^{(2)})^2 =d\, S\,\Omega^{(3)}_{0}(\Gamma)= d\, A^{(3)}_1
\eea
which implies the local triviality relation (\ref{localtrivialitytwo}). 

The gravitational CS action, however, is not a form and cannot
be integrated on a non-trivial manifold $M_3$. Thus the anomaly cannot
be removed, since it is not the variation of a genuine 3-form. 
An alternative way to state this  is to remark that 
the generalized anomaly form ${\mathcal A}_4$ is not $\delta$-exact, even locally. Indeed,  by defining the gravitational CS generalized form of total degree 3
\bea
&&\Omega_{3}(\Gamma) \equiv \Omega^{(3)}_{0}+\Omega^{(2)}_{1} +\Omega_2^{(1)}\nn\\
&&\Omega_2^{(1)}\equiv\frac{1}{ 2}{\rm tr}\,\Gamma\, R^{(0)}
\eea
one can rewrite relations (\ref{localtrivialityone}) and (\ref{localtrivialitytwo}) 
as follows
\be
{\cal A}_{4}= \delta\,\Omega_{3}(\Gamma) + \omega_{4}
\label{localtriviality}
\ee
$\omega_4$ is the generalized form of total fermionic degree 4
\bea
\omega_4= \frac{1}{ 2} {\rm tr}\, \bigl[
\partial\,\gamma\,\bigl(d\,\Gamma +R^{(1)}+R^{(0)}\bigr)\bigr]
\eea
where $\partial\, \gamma$ is the matrix $\partial_\mu\,\gamma^\nu$.
The component of form degree 3 of $\omega_4$ 
vanishes, but the components of lower degree do not. 
Thus, although the 3-form
anomaly is locally exact, the rest of the anomaly multiplet is not exact, 
even locally.

Note that Eq. (\ref{localtriviality}) is consistent with both  $\delta$-closeness of  ${\mathcal A}_4$  and a non-vanishing $\omega_4$ 
\be
\delta\,{\cal A}_{4}= 0= \delta^2\Omega_{3}(\Gamma) + \delta\,\omega_{4}
\label{localtrivialcocycleconsistency}
\ee
since $\delta$ is nilpotent only on forms 
\be
\delta^2 = {\cal L}_\gamma - \{i_\gamma, d\}
\ee
and $\Omega_{3}(\Gamma)$ is not a form.

\sectiono{Topological Anomaly Inflow}
\label{topinflow}
 
In this Section we discuss  possible implications 
of CS topological anomalies of higher ghost number 
for topological strings.

CS theory with $G=SU(N)$ on a 3-manifold $M_3$ describes a stack of $N$ topological D-branes  --- of the A-type --- propagating on the non-compact $CY$ 6-manifold  $X_6=T^* M_3$\cite{Witten:1992fb}.   Such an open string interpretation of CS theory led Witten to suggest that the  framing anomaly be cancelled by couplings coming from the closed string sector. Evidence for the existence of closed string couplings of this kind of  (more or less) the right magnitude was presented in \cite{Gopakumar:1998ki}.
In this Section we will try to extend these considerations to the higher-ghost deformation of $SU(N)$ CS theory that was discussed in Section \ref{higherghostCS}. We will work out the general form of the anomalous couplings
of the topological closed string field theory which  cancel all the higher ghost
CS topological anomalies (\ref{higherghostanomalies}).

The quantum properties of the target space  field theory describing closed  topological  strings of the  A-type are poorly understood, although a proposal for the
classical theory has been put forward in \cite{Bershadsky:1994sr}. 
The physical vertex operators of the closed A topological model 
are de Rahm cohomology classes of forms on $X_6$ of degree 2. Consequently 
it is natural to assume that the  corresponding target space theory contains a 2-form field $k^{(2)}$ whose linearized equations of motion read
\be
d\, k^{(2)}=0
\label{closedem}
\ee
and whose gauge properties are described by a BRST operator $S_0$
\bea
&& S_0\, k^{(2)} = d\, k^{(1)}\qquad S_0\, k^{(1)} = d\, k^{(0)}\qquad S_0\, k^{(0)} = 0
\eea
Repeating the arguments that we applied to CS theory,  one arrives to the conclusion that  the closed target space theory  must be coupled to {\it topological} 6-dimensional gravity by  introducing the equivariant BRST operator
\be
s = S - {\cal L}_\xi
\ee
where
\bea
&& S\, k^{(2)} = d\, k^{(1)}\qquad S\, k^{(1)} = d\, k^{(0)}-i_\gamma(k^{(2)})
\qquad S\, k^{(0)} = -i_\gamma(k^{(1)})\nonumber
\eea 
and $\gamma^\mu$ is the 6-dimensional reparametrizations super-ghost.
However,  $s$ so defined is nilpotent only up to the equations of motion of $k^{(2)}$:
\be
S^2 = {\cal L}_\gamma + i_\gamma(d\, k^{(2)}) \frac{\delta}{ \delta\, k^{(2)}}
\ee
We already know a way to fix this: Introduce form fields of higher degrees and define 
the generalized form
\be
{\cal K} =  k^{(0)}+k^{(1)}+k^{(2)}+k^{(3)}+k^{(4)}+k^{(5)}+k^{(6)}
\ee
The BRST transformations of the closed string fields $k^{(p)}$ are captured by
a coboundary operator $\delta$ 
\be
\delta\, {\cal K} \equiv  \bigl(S\, -d\ +i_\gamma\bigr)\, {\cal K}=0
\label{closedbrst}
\ee
which is nilpotent off-shell
\be
 \delta^2=0
\ee
The proposal in \cite{Bershadsky:1994sr} for the target space theory  
of A-type topological strings, advocates in fact, partly because of the CS analogy,  the introduction of string fields
$k^{(p)}$ with all $p$.  More precisely, forms $k^{(p)}$ with $p=0,1,2$ are identified in \cite{Bershadsky:1994sr} with {\it fields} while those with $p=3,4,5$ are interpreted
as {\it anti-fields}. There is a further constraint on the theory which leads to vanishing
6-form field $k^{(6)}$.  As we elaborate in the following, topological anomalies
considerations represent an independent reasoning that supports such a setting
for the closed target space theory.

Topological $D$-branes should act as sources for the closed string fields.
We expect therefore that in presence of branes wrapped around a (Lagrangian) 
3-cycle $M_3$ of $X_6$ the linearized equations of motion of the closed string
field (\ref{closedem}) acquire a source term
\be
d\, k^{(2)} = \alpha\, \delta_{M_3}
\ee
where $\alpha$ is a constant proportional to the D-brane charge and 
$ \delta_{M_3}$, the Poincar\'e dual of the cycle $M_3$,  is a closed 3-form with support on the brane.  Correspondingly,
the BRST transformations (\ref{closedbrst})   should
be modified,  in presence of branes, as follows
\be
\delta\, {\cal K} = \alpha\, \delta_{M_3}
\ee
Consistency requires
\be
\delta^2\, {\cal K}= \alpha\,\delta\, \delta_{M_3}=0
\ee
which is satisfied since
\be
d\, \delta_{M_3}=0\qquad i_\gamma\,\bigl(\delta_{M_3}\bigr) =0 \qquad S\,\delta_{M_3}=0
\ee
The first of these conditions is the closeness of the Poincar\'e dual of the cycle $M_3$.
The second equation is the statement that, in presence of branes, the diffeomorphisms which represent genuine gauge symmetries of the theory are those leaving
$M_3$ invariant. Finally, the last equation is the condition that $M_3$ be a supersymmetric cycle, which, for the A-model, is the requirement that $M_3$ be a Lagrangian sub-manifold of $X_6$ \cite{Witten:1992fb}.

In this setting there is a natural and simple  way to introduce the couplings of the
closed string theory which cancel the topological anomalies of the D-brane theory.
Each CS anomaly  with coefficient $c_{2\,(p-1)}(t_i)$ 
is cancelled by a term in the closed string field  action which couples
the closed string field to the topological gravity observable ${\cal A}_{4\,p}$:
\be 
I_p = -\frac{c_{2\,(p-1)}(t)}{\alpha}\int_{X_6} \, {\cal A}_{4\,p}\,{\cal K}=
-\frac{c_{2\,(p-1)}(t)}{\alpha}\int_{X_6} \bigl({\rm tr} {\cal R}^2\bigr)^{p}\, {\cal K}
\label{anomalouscouplings}
\ee
so that
\bea
&&S\, I_p = -\frac{c_{2\,(p-1)}(t)}{\alpha}\int_{X_6} {\cal A}_{4\,p}\,\delta\,{\cal K}=-c_{2\,(p-1)}(t)\int_{X_6} {\cal A}_{4\,p}\,\delta_{M_3}=\nn\\
&&=-c_{2\,(p-1)}(t) \int_{M_3}{\cal A}_{4\,p}= -S\,  F_{4\,(p-1)}
\eea
As mentioned earlier, arguments confirming the presence in the topological 
closed theory of the term $ I_1$ which cancels the $p=1$ framing anomaly 
have been presented 
in \cite{Gopakumar:1998ki}. Here we see that this could be just the tip
of an iceberg: To each one of the anomalies of the deformed $SU(N)$ CS
theory there should correspond, on the closed string side, a ``dressing''
of the closed string field by an appropriate topological gravity observable.

\section*{Acknowledgements}

I would like to thank M. Bonini, S. Cecotti, S. Giusto and, especially,  R. Stora for 
useful discussions. 
I am also grateful to the Theory Group of CERN, Geneva, Switzerland, 
for providing me with hospitality, support and a congenial working environment
during the 2008 Institute ``Black Holes, a landscape of
theoretical physics problems'',  when part of
this work was done. 

\appendix


\sectiono{Equivariant Cohomology} 
\label{equivariantcohomology}

In this Appendix we review some standard material about the equivariant cohomology
of topological gravity.

We will be interested in studying cohomologies of $s$ modulo $d$, the exterior
differential acting on space-time forms. It is convenient to introduce a fermionic degree $f$ which is the sum
of the ghost degree and the form degree.  The operators $S_0$, $S$, $d$ and $i_\gamma$ all have $f=+1$. It is therefore coherent to take both $d$ and $i_\gamma$ to {\it anti-commute} with $S_0$ and $S$.  With this choice 
the operator
\be
\tilde{\delta}\equiv s -d
\ee
is nilpotent when acting on generalized forms
\be
\tilde{O} \equiv \tilde{O}^{(0)}+\tilde{O}^{(1)}+\tilde{O}^{(2)}+\tilde{O}^{(3)}
\ee
$\tilde{\delta}$-cocycles
\be
\tilde{\delta}\, \tilde{O} =0
\label{scocycle}
\ee
are associated to the descent equations
\be
s\, \tilde{O}^{(3)}= d\, \tilde{O}^{(2)}\quad s\, \tilde{O}^{(2)}= d\, \tilde{O}^{(1)}\quad s\, \tilde{O}^{(1)}= d\, \tilde{O}^{(0)}\quad 
s\, \tilde{O}^{(0)}= 0
\label{sdescent}
\ee
and trivial $\tilde{\delta}$-cocycles 
\be
\tilde{O}= \tilde{\delta}\, \tilde{\omega}
\label{trivialcocycle}
\ee
correspond to
\bea
&&\tilde{O}^{(3)}= s\, \tilde{\omega}^{(3)}-d\, \tilde{\omega}^{(2)}\quad\tilde{O}^{(2)}= s\,\tilde{\omega}^{(2)}- d\, \tilde{\omega}^{(1)}\nonumber\\
&&\tilde{O}^{(1)}= s\, \tilde{\omega}^{(1)}-d\, \tilde{\omega}^{(0)}\quad
\tilde{O}^{(0)}= s\, \tilde{\omega}^{(0)}
\label{trivialdescent}
\eea
Topological gravity instructs us to consider a particular $s$ cohomology,
the {\it equivariant} cohomology.  This means considering not just any
$\tilde{\delta}$-cocycles but those whose dependence on the reparametrization
ghost is restricted to be of the following form
\be
\tilde{O} = {\rm e}^{i_\xi}\, O
\label{equivariantdeltacocycle}
\ee
where 
\be
O \equiv O^{(0)}+O^{(1)}+O^{(2)}+O^{(3)}
\ee
is a generalized form which does {\it not} depend on $\xi$. 
Writing (\ref{equivariantdeltacocycle}) in terms of the components of 
fixed form degree one has
\bea
&&\tilde{O}^{(3)}= O^{(3)}\nonumber\\
&&\tilde{O}^{(2)}= O^{(2)}+i_\xi\,\bigl(O^{(3)}\bigr)\nonumber\\
&&\tilde{O}^{(1)}= O^{(1)}+i_\xi\,\bigl(O^{(2)}\bigr)+
\frac{1}{ 2}\, i_\xi\,i_\xi\,\bigl(O^{(3)}\bigr)\nonumber\\
&& \tilde{O}^{(0)}= O^{(0)}+ i_\xi\,\bigl(O^{(1)}\bigr)+\frac{1}{ 2}\, i_\xi\,i_\xi\,\bigl(O^{(2)}\bigr)+\frac{1}{ 3!}\, i_\xi\,i_\xi\,i_\xi\,\bigl(O^{(3)}\bigr)
\label{equivariantcocycle}
\eea
where $O^{(k)}$, with $k=1,2,3$, are $\xi$-independent.

The following coboundary operator can be defined on the space of $\xi$-independent generalized forms
\be
\delta \equiv S+i_\gamma- d
\label{equivariantdelta}
\ee
$\delta$ is nilpotent thanks to the relations
\be
S^2 ={\cal L}_\gamma = \{ d, i_\gamma\}
\ee
It is easily shown that equivariant $\tilde{\delta}$-cocycles 
$\tilde{O} = {\rm e}^{i_\xi}\, O$  are in
one-to-one correspondence with ($\xi$-independent) $\delta$-cocycles $O$.
The  components of fixed form degree of $\delta$-cocycles
are in the cohomology of $S$ modulo $d$ and modulo $i_\gamma$:
\bea
&&S\,O^{(3)}= d\, O^{(2)}\nonumber\\
&& S\,O^{(2)}= d\,O^{(1)}- i_\gamma\bigl(O^{(3)}\bigr)\nonumber\\
&& S\,O^{(1)}= d\,O^{(0)}- i_\gamma\bigl(O^{(2)}\bigr)\nonumber\\
&& S\,O^{(0)}= -i_\gamma\bigl(O^{(1)}\bigr)
\label{sgammadescent}
\eea
Summarizing, the physically relevant equivariant $s$-cohomology
modulo $d$ is in a  one-to-one correspondence with the $S$-cohomology modulo
$d$ and modulo $i_\gamma$ on the space of $\xi$-independent operators.

The relation between the $S$-cohomology (modulo $d$ and modulo $i_\gamma$) 
and  the ``naive'' $S_0$-cohomology (modulo $d$) is based on
the decomposition
\be
S= S_0 +G_\gamma
\ee
which leads to 
\be
\delta \equiv \delta_0 + \delta_\gamma
\ee
where
\be
\delta_0 \equiv S_0 -d\qquad \delta_\gamma = G_\gamma +i_\gamma
\ee
$\delta_0$ and $\delta_\gamma$ are  (anti)-commuting coboundary operators
\be
\delta_0^2 =\delta_\gamma^2 = \{\delta_\gamma,\delta_0\}=0
\ee

Since $\delta_0$ and $\delta_\gamma$ (anti)-commute, we can consider
the cohomology of $\delta_0$ {\it relative} to $\delta_\gamma$. This is
defined on the $\delta_\gamma$-invariant subspace of ``matter'' operators --- 
i.e. those  independent of the gravitational fields.
In the context of $N=2$ supersymmetric field
theories these operators are called ``chiral''.
The cohomology of $\delta_0$ relative to $\delta_\gamma$ is the kernel
of $\delta_0$ acting on ``chiral'' operators, modulo $\delta_0$-trivial
operators $\delta_0\,\omega$ with $\omega$ ``chiral''.
 
A matter operator which is both ``chiral'' and  $\delta_0$-closed --- i.e.  a 
class in the relative $\delta_0$ cohomology ---  maps, by means of the
identity map, to a class of the equivariant 
$\delta$. In Section \ref{higherghostCS} we consider such kind of
observables. 

It should be emphasized that the  map between relative 
$\delta_0$-cohomology and equivariant cohomology is not, in general, one-to-one:
both its surjectiveness and injectiveness depend on the cohomology 
of $\delta_\gamma$. Only if $\delta_\gamma$ had no
cohomology, equivariant $\delta$-cohomology and the relative 
$\delta_0$ cohomology would be the same.

\sectiono{A Cohomological Identity} 
\label{filtration}

We want to show that the differences
\bea
&&{\mathcal B}_{2\,k} \equiv {\rm tr}\,\hat{\mathcal R}^{k}-{\rm tr}\,{\mathcal R}^{k} 
= \nn\\
&&\qquad = {\rm tr}\,{\mathcal F}^k -\,k\, {\rm tr}\,{\mathcal F}^{k-1}\, {\mathcal R}+\cdots
\eea
are $\delta$-trivial. Let us start with a more general preliminary proposition. 
Consider an invariant polynomial $P_k({\mathcal F},\hat{\mathcal R}, g)$  of total fermionic degree $2\,k$ built with the matrices ${\mathcal F}^\mu_{\nu}$,
 $\hat{\mathcal R}^\mu_\nu$, and  $g^{\mu\nu}$ . 
We will show that if $P_k({\mathcal F},\hat{\mathcal R},g)$
is $\delta$-closed, then it is cohomologically equivalent to a class which
does not contain ${\mathcal F}$.

Let us introduce 
a grading which counts the numbers of ${\mathcal F}$.
We are going to decompose $P_k({\mathcal F},\hat{\mathcal R},g)$  into
components $X_i({\mathcal F},\hat{\mathcal R},g)$ of fixed 
${\mathcal F}$-degree $i$, with $i=0,1,\ldots k$:
\be
P_k({\mathcal F},\hat{\mathcal R},g) = X_k({\mathcal F},\hat{\mathcal R},g) + X_{k-1}({\mathcal F},\hat{\mathcal R},g) +\cdots +X_0(\hat{\mathcal R},g)
\label{Fdecomposition}
\ee

Recall the action of the $\hat{\delta}$-differential (\ref{deltahatX})
on the algebra spanned
by  ${\mathcal C}$, ${\mathcal F}$ and $\hat{\mathcal R}$:
\bea
&&\hat{\delta}\, {\mathcal C} = {\mathcal F} + {\mathcal C}^2\nn\\
&&\hat{\delta}\, {\mathcal F} = \bigl[{\mathcal F},{\mathcal C}\bigr]+
\bigl[\hat{\mathcal R},{\mathcal C}\bigr]\nn\\
&&\hat{\delta}\, \hat{\mathcal R} = 0
\eea
Moreover
\be
\hat{\delta}\, g^{\mu\nu} = -\psi^{\mu\nu}+\frac{1}{2}\,\psi^\mu_\lambda\,g^{\lambda\nu}+\frac{1}{2}\,\psi^\nu_\lambda\,g^{\lambda\mu}=0
\ee
Let us decompose $\hat{\delta}$ according to its ${\mathcal F}$-grading:
\be
\hat{\delta} = \hat{\delta}_1 + \hat{\delta}_0+ \hat{\delta}_{-1}
\ee
$\hat{\delta}_m$ for $m=-1, 0,1$ have  ${\mathcal F}$-degree  equal to  $m$:
\bea
&&\hat{\delta}_1\,{\mathcal C} = {\mathcal F}\qquad \hat{\delta}_1\,{\mathcal F} =0\nn\\
&&\hat{\delta}_0\,{\mathcal C} = {\mathcal C}^2\qquad \hat{\delta}_0\,{\mathcal F} =\bigl[{\mathcal F},{\mathcal C}\bigr]\qquad 
\nn\\
&&\hat{\delta}_{-1}\,{\mathcal C} = 0\qquad \hat{\delta}_{-1}\,{\mathcal F} =\bigl[\hat{\mathcal R},{\mathcal C}\bigr]\qquad \nn\\
&& \hat{\delta}_{m}\,\hat{\mathcal R} =\hat{\delta}_m\, g=0\qquad {\rm for}\; m =-1,0,1
\eea
One verifies that
\bea
&&\hat{\delta}_1^2 = \hat{\delta}_0^2=\hat{\delta}_{-1}^2= \{\hat{\delta}_0,\hat{\delta}_{1}\}=\{\hat{\delta}_0,\hat{\delta}_{-1}\}=0
\eea
and
\bea
&&\{\hat{\delta}_1,\hat{\delta}_{-1}\}\,X = \bigl[\hat{\mathcal R}, X\bigr]
\eea
on any matrix-valued generalized form $X$. 

Decomposing the equation
\be
\hat{\delta}\,P_k({\mathcal F},\hat{\mathcal R},g)=0
\ee
in components of fixed ${\mathcal F}$-degree one derives the descent equations
\bea
&& \hat{\delta}_1\,X_k=0\nn\\
&&  \hat{\delta}_1\,X_{k-1}+\hat{\delta}_0\,X_k=0\nn\\
&&\hat{\delta}_{1}\,X_{k-2}+\hat{\delta}_0\,X_{k-1}+\hat{\delta}_{-1}\,X_{k}=0\nn\\
&& \cdots\nn\\
&&\hat{\delta}_1\,X_0+\hat{\delta}_0\,X_{1}+\hat{\delta}_{-1}\,X_{2}=0\nn\\
&&\qquad\quad\; \hat{\delta}_0\,X_{0}+\hat{\delta}_{-1}\,X_{1}=0\nn\\
&&\qquad \qquad\qquad\; \hat{\delta}_{-1}\,X_{0}=0
\label{filterdescent}
\eea
The basic observation is that the nilpotent $\hat{\delta}_1$ has, obviously,  vanishing
cohomology on sectors with non-zero ${\mathcal F}$-degree. Therefore, from the first of the equations (\ref{filterdescent}) one deduces
\bea
&&X_k =\hat{\delta}_1\,Y_{k-1}
\label{firstfilter}
\eea
which inserted in the second equation gives
\bea
&&  \hat{\delta}_1\,\bigl(X_{k-1}-\hat{\delta}_0\,Y_{k-1}\bigr)=0\nn
\eea
This in turns implies
\bea
&&X_{k-1} =\hat{\delta}_1\,Y_{k-2}+\hat{\delta}_0\,Y_{k-1}\nn\\
&&X_{k-2} =\hat{\delta}_1\,Y_{k-3}+\hat{\delta}_0\,Y_{k-2} +\hat{\delta}_{-1}\,Y_{k-1}
\label{secondfilter}
\eea
and so on, until one reaches the  equation:
\bea
X_0 =\hat{\delta}_0\,Y_{0}+\hat{\delta}_{-1}\,Y_{1} +Z(\hat{\mathcal R},g)
\label{lastfilter}
\eea
where $Z(\hat{\mathcal R},g)$ does not depend on ${\mathcal F}$.  Putting 
(\ref{lastfilter}) together with the previous equations  one finally obtains
\bea
P_k({\mathcal F},\hat{\mathcal R},g)=\hat{\delta}\, \bigl(Y_0+Y_1+\cdots+Y_{k-1}\bigr)+ Z(\hat{\mathcal R},g)
\eea
which is our preliminary proposition. Let us apply it to ${\mathcal B}_{2\,k}$.
In this case the component of zero ${\mathcal F}$-degree vanishes
\be
X_0 =0 = \hat{\delta}_0\,Y_{0}+\hat{\delta}_{-1}\,Y_{1} +Z(\hat{\mathcal R},g)
\ee
But both $\hat{\delta}_{0}$ and $\hat{\delta}_{-1}$ increase the number of ${\mathcal C}$ by one. Therefore
\be
 \hat{\delta}_0\,Y_{0}+\hat{\delta}_{-1}\,Y_{1}=0
\ee
and
\be
Z(\hat{\mathcal R},g) =0
\ee
which is what we wanted to show.

\end{document}